\documentclass[a4paper,twocolumn,pra,superscriptaddress,amsmath,amssymb,mathrsfs,showpacs]{revtex4-1}
\usepackage{graphicx}
\usepackage{epstopdf}
\usepackage[table]{xcolor}     
\usepackage{braket}
\usepackage[colorlinks]{hyperref} 

\hypersetup{%
linkcolor=blue,
citecolor=blue,
filecolor=black,
menucolor=black,
urlcolor=black
}
\usepackage[customcolors,norndcorners]{hf-tikz}
\hfsetfillcolor{gray!10}
\hfsetbordercolor{gray!50}

\begin{document} 
\title{An adaptive-size multi-domain pseudospectral approach for solving  the time-dependent Schr\"odinger equation}     
\author{R. Esteban Goetz}
\affiliation{Theoretische Physik, Universit\"at Kassel,\ Heinrich Plett-Stra\ss e 40,\ 34132 
Kassel, Germany}
\author{Andrea Simoni}
\affiliation{Institut de Physique de Rennes, UMR 6251, CNRS and Universit\'e de Rennes 1, 35042 Rennes Cedex, France}
\author{Christiane P. Koch}
\email{christiane.koch@uni-kassel.de}
\affiliation{Theoretische Physik, Universit\"at Kassel,\ Heinrich
  Plett-Stra\ss e 40,\ 34132  
  Kassel, Germany}
\date{\today}
\begin{abstract}
  We show that a pseudospectral representation of the wavefunction using multiple spatial domains of variable size yields a highly accurate, yet efficient method to solve the time-dependent Schr\"odinger equation. The overall spatial domain 
is split into non-overlapping intervals whose size is chosen according to the local de Broglie wavelength.  A multi-domain weak formulation of the Schr\"odinger equation is obtained by representing the wavefunction by  Lagrange polynomials with compact support in each domain, discretized at the  Legendre-Gauss-Lobatto points. The resulting Hamiltonian is sparse, allowing for efficient diagonalization and storage. Accurate time evolution is carried out by the Chebychev propagator, involving only sparse matrix-vector multiplications. Our approach combines the efficiency of mapped grid methods with  the accuracy of spectral representations based on Gaussian quadrature rules and the stability  and convergence properties of polynomial propagators. 
We apply this method to high-harmonic generation and examine the role of the initial state for the harmonic yield near the cutoff. 
\end{abstract}

\maketitle

\section{Introduction}

Over the past decades, the field of quantum molecular dynamics has been driven  by the development of efficient numerical methods for solving the time-dependent
Schr\"odinger equation~\cite{TannorBook}. Current applications include studies of quantum optimal control~\cite{GlaserEPJD15} or electron dynamics. 
The two basic tasks that need to be addressed in quantum molecular dynamics are the representation of the state vector (and operators acting on it) and its time evolution. Ideally, the accuracy with which both tasks are accomplished should be balanced~\cite{KosloffJPC88}. High accuracy is required by many 
state-of-the-art applications, for example in quantum optimal control~\cite{GlaserEPJD15}. 
At the same time, the exponential scaling of quantum dynamics calculations is a challenge for even the most advanced computer architectures.  Efficiency of the methods is therefore also an issue, in addition to accuracy. 

Highly accurate methods are obtained by employing pseudospectral approaches based on the expansion in orthogonal polynomials~\cite{boyd}. The representation problem can be solved using discrete variable representations or their unitary equivalent, finite basis representations~\cite{dvr}. 
The operators acting on the wavefunction are then given as sparse (often diagonal) matrices in one of the representations. The numerical effort is either due to the unitary transformation connecting the two representations or due to sparse matrix-vector multiplications. For a sufficiently large number of basis functions, the error becomes smaller than machine precision~\cite{RonnieGrid96}. 

Polynomial approximations yield also the most accurate and stable propagation schemes~\cite{RonnieReview94}. Again, convergence is exponential with increasing  polynomial order. For coherent time evolution, the best polynomial approximation of the evolution operator is obtained by the 
Chebyshev propagator~\cite{TalEzerJCP84},  while Newton polynomials yield an accurate and efficient propagator for open quantum systems~\cite{RonnieReview94}. Modifications of polynomial propagators allow to also accurately account for time ordering in case of a time-dependent Hamiltonian~\cite{PeskinJChemPhys93,NdongJCP10,Tal-EzerJSC12}.

The high accuracy of these methods may, however, be compromised in time-dependent studies of dissociation or ionization where a sufficiently large grid, respectively a sufficiently large number of basis functions, becomes computationally prohibitive, both in terms of storage requirements and CPU time. Remedies to this problem  include the use of variable grid steps~\cite{FattalPRE96,KokoulineJCP99,WillnerJCP04,KallushCPL06} or variable-grid boundary conditions~\cite{NurhudaPRA99}, wavefunction splitting methods~\cite{HeatherMetiu,KellerPRA95,AntoniaPRA14},  mask functions~\cite{KrausePRA92,ChelkowskiPRA95,KulanderJOptB90} or 
complex absorbing potentials (CAPs)~\cite{SantraPhysRep02,MugaPhysRep04}.
While the latter approach allows for calculating physical observables
that require long propagation times~\cite{GreenmanPRA10,RohringerPRA09,WoppererThesis2013}, 
a CAP can only absorb wavepacket components within a certain frequency range~\cite{MugaPhysRep04}. It is thus rather difficult to completely avoid reflection which compromises accuracy. The problem of reflection also occurs for the mask function approach~\cite{NurhudaPRA99}.
A CAP, moreover, renders the
Hamiltonian non-Hermitian, which results in substantial 
technical difficulties for quantum optimal control~\cite{GreenmanPRA15,OhtsukiJChemphys99,GoetzPRA16} and may even preclude  the evaluation of observables of interest~\cite{RohringerPRA09,PabstPRL2011,GreenmanPRA10}.
Non-Hermitian Hamiltonians are avoided when using variable grid steps or wavefunction splitting but also in these cases high accuracy and reasonable numerical costs are not always guaranteed. For example, the mapped Fourier grid method~\cite{KokoulineJCP99,WillnerJCP04,KallushCPL06,TiesingaPRA98} was developed for long-range potentials that vanish asymptotically as $1/R^N$. It allows for an accurate description of most bound states and low-energy scattering states. However, the calculation of the bound spectrum does not scale favorable with the number of grid points, rendering its application in coupled channel calculations difficult~\cite{GonzalezPRA12,CrubellierNJP15}. Moreover, it cannot be used in photoionization studies where high-energy scattering states may come into play. Wavefunction splitting is applicable in this case~\cite{AntoniaPRA14,GoetzPRA16}; it neglects, however, the Coulomb interaction between photoion and photoelectron. Thus, it cannot be used to study processes where recombination of the photoelectron is crucial, such as high harmonic generation. 
Here, we use a weak formulation~\cite{Chipot2009,boyd} of the Schr\"odinger equation to derive a sparse, yet accurate representation of the Hamiltonian and combine it with the Chebyshev propagation method~\cite{TalEzerJCP84}. 
The basic idea is to decompose the spatial domain into multiple sub-intervals of increasing size, chosen according to the local de Broglie wavelength, similar to the choice of the variable grid step in the mapped Fourier grid method~\cite{KokoulineJCP99,WillnerJCP04,KallushCPL06,TiesingaPRA98}. Within each interval, the wavefunction is expanded into Lagrange polynomials. The resulting representation of the Hamiltonian is sparse which is exploited in storage, diagonalization and matrix-vector operations~\cite{Lehoucq97arpackusers}. Our approach thus combines the high accuracy of pseudospectral methods with the ability to use a very large spatial domain. It is particularly advantageous for quantum dynamics involving long-range potentials and 
long propagation times. As an example, we consider a laser-driven electron in a soft Coulomb potential, a popular model for high-harmonic generation. 

The paper is organized as follows. Section~\ref{sec:method} presents the method, starting with a brief review of the Chebychev propagator. The pseudospectral multi-domain representation of the Hamiltonian is derived from the weak formulation of the Schr\"odinger equation in Section~\ref{subsec:weak}. Within each domain, a Gauss-Lobatto-Legendre collocation is employed, as described in Section~\ref{subsec:colloc}, and a global representation of the Hamiltonian is derived in Section~\ref{subsec:global} by assembling all domains.
In section III, we present and compare numerical results between the MFGH and the MFGH-SEM. Section IV is 
devoted to time dependent results. Finally, summarizing remarks are 
outlined in Section V.  

\section{Method}
\label{sec:method}

The time-dependent Schr\"odinger equation reads
\begin{equation}
  \label{eq:tdse}
   i\hbar\dfrac{\partial}{\partial t}\Psi(r,t) = \hat{H}\Psi(r,t)\,,   
\end{equation}
where the Hamiltonian, 
\begin{equation}
  \label{eq:hamiltonian:def}
  \hat{H} =-\dfrac{\hbar^2}{2m}\nabla^2 + V(r,t)   \,,
\end{equation}
may contain a time-dependent term. 
The formal solution is given by 
\begin{eqnarray}
  \Psi(r,t) &=& \hat{T} \exp\left(-\dfrac{i}{\hbar}\int_0^t
       \hat H(\tau)d\tau\right)\Psi(r,0)\nonumber\\
       &=&  \hat{U}(t)\Psi(r,0)\,.
\end{eqnarray}
Polynomial propagators expand the evolution operator, $\hat{U}(t)$, as a function of the Hamiltonian, in a truncated polynomial series~\cite{TalEzerJCP84,leforestierJCP91,NdongJCP10}. To this end, the domain of the eigenvalues, i.e., the spectral radius of the Hamiltonian $\Delta E$, must be known. Consider the example of a time-independent Hamiltonian in which case the Chebychev propagator is simply obtained as  
\begin{eqnarray}
\label{eq50}
\hat{U}(t) = e^{-i\hat{H}t/\hbar}\approx\sum_{n=0}^{N}a_n T_n\left(-i\hat{H}t/\hbar\right)\, .
\end{eqnarray}
Since the complex Chebyshev polynomials are defined in the interval
$[-i,i]$, the Hamiltonian must be renormalized,
\[
\hat H_{norm}=\frac{2\left(\hat H-\openone\left(\Delta E/2+V_{min}\right)\right)}{\Delta E}\,.
\]
The expansion coefficients $a_n$ are known analytically~\cite{TalEzerJCP84} and the  Chebyshev polynomials can be computed using their recursion formula. The solution is thus obtained by subsequent applications of the (renormalized) Hamiltonian to a wavefunction~\cite{TalEzerJCP84,KosloffJPC88,RonnieReview94}. 
For a prespecified error, the number of Chebyshev polynomials, i.e., the number of times the Hamiltonian is applied to a wavefunction, is determined by the product of spectral radius $\Delta E$ and time step~\cite{TalEzerJCP84,KosloffJPC88,RonnieReview94}. 
If the Hamiltonian has a matrix representation, the propagation involves a series of matrix-vector multiplications, 
$\hat{H}_{norm}\Psi_n(r,0)$. 

Here, we derive a sparse representation of the
Hamiltonian~\eqref{eq:hamiltonian:def}. It is based on 
domain decomposition~\cite{boyd,QuarteroniBook,ToselliBook},
the variational or weak solution~\cite{Chipot2009} of the Schr\"odinger equation, 
Lagrange interpolation~\cite{boyd,Nevai1984}, and polynomial series
expansions of operators~\cite{TalEzerJCP84,KosloffJPC88,RonnieReview94}. The concept of the local de Broglie wavelength, central to the mapped Fourier grid Hamiltonian~\cite{KokoulineJCP99,WillnerJCP04,KallushCPL06,TiesingaPRA98}, is used to determine the size of the domains. 

\subsection{Multi-domain weak formulation}
\label{subsec:weak}
In order to derive a matrix representation of the Hamiltonian~\eqref{eq:hamiltonian:def}, we consider the time-independent radial Schr\"odinger equation, 
\begin{eqnarray}
\label{eq:tise}
-\dfrac{\hbar^2}{2\mu}\nabla^2 u(r) + V(r)u(r) = \lambda\, u(r)
\end{eqnarray}
with $r\in\Omega=[r_{min},r_{max}]$ and $\lambda$ an eigenvalue. We employ domain decomposition for $\Omega$. The main idea behind this method is to split the domain of (spatial) integration $\Omega$ into $M$ non-overlapping intervals, or 'elements', $\Omega_k$ of arbitrary size. The total domain, $\Omega$, is constructed from the union of the $M$ non-overlapping elements,
\begin{eqnarray}\label{eq4}
 \Omega = \bigcup_{k = 1}^M \Omega_k\mbox{ with }
\Omega_k \cap\Omega_{k^\prime} 
 =
  \begin{cases}
    \{r^k_N\} &\mbox{if } k^\prime=k+1,\\
    \varnothing &\mbox{otherwise},
  \end{cases}
\end{eqnarray}
where $ r^k_N = r^{k+1}_0$ and each interval $\Omega_k=[r^k_0, r^k_N]$ will be discretized using $N+1$ points, 
and the constraint $r^k_N = r^{k+1}_0$ ensures connection of all $\Omega_k$.
Within each interval $\Omega_k\in\Omega$, Eq.~\eqref{eq:tise} becomes
\begin{eqnarray}
\label{eq23}
-\dfrac{\hbar^2}{2\mu}\nabla^2u^k(r) + V(r)u^k(r)=\lambda\, u^k(r)\, . 
\end{eqnarray}
with $r\in\Omega_k$. In order to derive the weak solution of the Schr\"odinger equation for a given $\Omega_k$,  we multiply both sides of Eq.~\eqref{eq23} by an arbitrary test function, $v^k(r)\in H^1(\Omega_k)$, where $H^1(\Omega^k)$ refers to the Sobolev space defined as
\begin{eqnarray}
\label{Sobolev}
H^1(\Omega^k) = \bigg\{ \phi\in L^2(\Omega^k)\, , \nabla \phi\in L^2(\Omega) \bigg\}\, . 
\end{eqnarray}
Integrating over the domain $\Omega_k$ and applying Green's theorem, we find 
\begin{widetext}
\begin{eqnarray}
\label{eq25}
-\frac{\hbar^2}{2\mu}
\int_{\Omega_k} \nabla v^k(r)\nabla u^k(r) dr + \int_{\Omega_k} v^k(r) V(r) u^k(r) dr
+ \frac{\hbar^2}{2\mu}\oint_{\partial\Omega_k} v^k(r) \nabla_{n} u^k(r)\, d \Gamma = \lambda \int_{\Omega_k} v^k(r) u^k(r)\, dr\,,
\end{eqnarray}  
\end{widetext}
where $\nabla$ denotes the usual gradient and $\nabla_n$ stands for the normal derivative.
The solution $u^k(r)$ satisfying  Eq.~\eqref{eq25} is called the \textit{weak solution} on $\Omega_k$, as
opposed to the strong solution, i.e., $u^k(r)$ satisfying Eq.~\eqref{eq23}.
Note that $u^k(x)\in H^1(\Omega_k)$. The weak solution of the Schr\"odinger equation in the weak formulation is obtained by determining $u^k(x)\in H^1(\Omega_k)$ and $\lambda$ such that
\begin{subequations}\label{eq:bilinear}
\begin{eqnarray}
\label{eq29}
\left\{
\begin{array}{lcl}
 a^k(u,v) &=& \lambda(u,v)_{\Omega_k}\vspace{0.25cm}\\
  u(r) &=& \tilde{u}(r)\hspace{1.0cm} \text{in}\hspace{1.0cm} \partial\Omega_k\,,
\end{array}
\right.
\end{eqnarray}
where $\tilde{u}(r)$ stands for the boundary condition of
$u^k$ at the domain boundary, $\partial\Omega_k$,
and the bilinear forms, $a^k(\cdot,\cdot)$ and $(\cdot,\cdot)_{\Omega_k}$ are defined as follows,
\begin{widetext}
\begin{eqnarray}
\label{eq30}
  a^k(u,v) &=&  \frac{\hbar^2}{2\mu}\int_{\Omega_k} \nabla v^k(r)\nabla u^k(r)dr
               +\int_{\Omega_k} v^k(r) V^k(r) u^k(r)\, dr
 + \frac{\hbar^2}{2\mu}\oint_{\partial\Omega_j} v^k(r) \nabla_{n} u^k(r)\, d\Gamma\,, \\
\label{eq31}
(u,v)_{\Omega_k} &=& \int_{\Omega_k}u^k(r)v^k(r)dr \,.
\end{eqnarray}  
\end{widetext}
\end{subequations}
In order to derive an explicit representation of the Hamiltonian from the weak formulation of the Schr\"odinger equation, we rewrite the bilinear forms as a linear operator equation in dual space. To this end, we employ a Galerkin-type method based on piecewise cardinal functions with bounded support in $\Omega_k$,
$\delta^k(r-r_j)$ where $r_j\in\Omega_k$.

\subsection{Gauss-Lobatto-Legendre collocation}
\label{subsec:colloc}

Consider the vector space spanned  by $N+1$ cardinal functions defined within
$\Omega_k$ and denote the set of basis functions by $\{v_j^k\}_{j=0,\ldots,N}$.
We can expand $u^k(r)$ in this basis, 
\begin{eqnarray}
  u^{k}(r) &=& \sum_{j=0}^{N} u^{k}(r_j)v^k_j(r)
 = \sum_{j=0}^{N}u^k(r_j)\delta^k(r-r_j)\, . 
\label{eq24}
\end{eqnarray}
Inserting Eq.~\eqref{eq24} into Eq.~\eqref{eq25}, multiplying both sides of Eq.~\eqref{eq25} by one of the cardinal functions and integrating over $\Omega_k$, we find a set of $N+1$ algebraic equations, 
\begin{eqnarray}
\label{eq32}
\sum_{j=0}^{N}u^k_j a^k(v_i,v_j) &=& \lambda\sum_{j=0}^{N}u^k_j\left(v_i,v_j \right)_{\Omega_k}\,,
\end{eqnarray}
where $i=0,\ldots,N$, $u^k_j=u^k(r_j)$ and $r_j\in\Omega_k$ by construction. 

In the particular case of a discrete variable representation~\cite{dvr,SzalayJCP93}, the expansion coefficients $u^{k}(r_j)$ in Eq.~\eqref{eq24} correspond to the wavefunction amplitudes at every collocation point and the error is only due to the Gaussian quadrature approximation. In other words, in each domain $\Omega_k$, 
$u^k(r)$ is approximated at the collocation points by the interpolant in Eq.~\eqref{eq24}. Correspondingly, we can evaluate the integrals in Eq.~\eqref{eq30} by means of a Gaussian quadrature rule within each interval $\Omega_k$, 
\begin{eqnarray}
\label{eq7}
\int_{\Omega_k}f(r)dr = \sum_{j=0}^N f^k(r_j) w^k_j\,.
\end{eqnarray}
This leads to
\begin{subequations}
\begin{widetext}
\begin{eqnarray}
\label{eq33}
a^k(v_i,v_j) &\approx& \frac{\hbar^2}{2\mu}\sum_{q=0}^N \nabla v^k_i(r_q)\nabla v^k_j(r_q)w_q^k + \sum_{j=0}^N v^k_i(r_q)V(r_q)v^k_j(r_q)w^k_q + 
\frac{\hbar^2}{2\mu}\left( \nabla u^k(r^k_0)\delta_{0,i} -\nabla u^k(r^k_N)\delta_{N,i}\right)\,,\quad\quad
\end{eqnarray}  
\end{widetext}
Using Gauss-Lobatto sampling points, i.e., sampling points that include the boundary of the domain $\Omega_k$, 
by definition $v^k_i(r^k_0)=\delta_{0,i}$ and $v^k_j(r^k_N)=\delta_{N,i} $ for 
$k = 2,\dots, M-1$, i.e., for all domains except those containing $r_{min}$ and $r_{max}$. Analogously, for Eq.~\eqref{eq31} we use 
the discrete inner product in $\Omega_k$ which is given by
\begin{eqnarray}
\label{eq34}
(v_i,v_j)_{\Omega_k} &\approx& \sum_{q=0}^N u^k_i(r_q)v^k_j(r_q)w_q^k
= w^k_i\delta_{i,j}\, .
\end{eqnarray}
\end{subequations}

We will employ Gaussian quadrature based on Legendre polynomials. Since Legendre polynomials are defined on the interval $\Lambda=[-1,1]$, we need an affine transformation, 
\begin{subequations}\label{eq:affine}
\begin{eqnarray}
\label{eq5}
\Phi^k &:& \Lambda  \longrightarrow \Omega_{k} \nonumber\\
 && \xi_i \longmapsto  
\xi_i\left(r_N^{k}- r_0^{k}\right)/2 + 
\left(r_N^{k} + r_0^{k}\right)/2\, .
\end{eqnarray}
with Jacobian 
\begin{eqnarray}\label{eq:Jac}
  \mathcal{J}_{k} &=& (r_N^k - r^k_0)/2
\end{eqnarray}
\end{subequations} 
and $\xi_i$ the standard Gauss-Lobatto-Legendre sampling points, 
cf. Eq.~\eqref{eq3}.
Integration in $\Omega_k$ can then be directly connected to integration in $\Lambda$, 
\begin{eqnarray*}
\int_{\Omega_k}f(r)dr &=& \int^{+1}_{-1}f\circ \Phi^k(\xi)\, \mathcal{J}_k \, d\xi =
\mathcal{J}_{k}\, \sum^N_{j=0} f(r_j)\, w^{\Lambda}_j\,. 
\end{eqnarray*}
Comparing this to Eq.~\eqref{eq7}, we find
\begin{eqnarray}
\label{eq9}
w^k_j = \mathcal{J}_{k}\,w^{\Lambda}_j
\end{eqnarray}
with $w^{\Lambda}_j$ that standard Legendre quadrature weights, cf. Eq.~\eqref{eq3}.

Next, we evaluate the derivatives in Eq.~\eqref{eq33} in terms of derivatives of the cardinal functions, 
\begin{eqnarray}
\label{eq12}
\dfrac{\partial}{\partial r}f^k(r) = \sum_{j=0}^{N}f(r_j)\dfrac{\partial}{\partial r}\delta^k(r-r_j)\, .
\end{eqnarray}
Using the properties of the Legendre polynomials and the cardinal functions, cf. Appendix~\ref{sec:Legendre}, 
the first order differentiation matrix for Legendre cardinal functions is found to read~\cite{boyd} 
\begin{equation}
\label{eq17}
\dfrac{\partial}{\partial\xi} \delta^{(\Lambda)}(\xi_i-\xi_j)  = 
\begin{cases}
  -N(N+1)/4 & \mbox{if } i=j=0 \\ 
  N(N+1)/4 & \mbox{if } i=j=N \\ 
  0 & \mbox{if } 1\le i=j\le N-1\\
  \dfrac{L_N(\xi_i)}{L_N(\xi_j)(\xi_i-\xi_j)} & \mbox{if } j\neq i\,.
\end{cases}
\end{equation}
This, together with the affine transformation~\eqref{eq:affine}, allows to determine the derivatives in Eq.~\eqref{eq33} with an error that is solely due to the Gaussian quadrature approximation.
Moreover, a useful expression to evaluate the first term in Eq.~\eqref{eq33}, needed to construct the matrix representation of the kinetic operator in the Schr\"odinger equation, is given by
\begin{eqnarray*}
S^k_{\mu,\nu} = \int_{\Omega_k} \dfrac{\partial }{\partial r}\delta^k(r-r_\mu) \dfrac{\partial }{\partial r}\delta^k(r-r_\nu)\, dr^k
\end{eqnarray*}
with $\mu,\nu = 0,\dots,N$. It is straightforward to show that
\begin{eqnarray*}
S^k_{\mu,\nu} =  \mathcal{J}^{-1}_k S^{\Lambda}_{\mu,\nu}\, ,
\end{eqnarray*}
where 
\begin{eqnarray}
\label{eq21}
S^{\Lambda}_{\mu,\nu} &=& \int_{\Lambda} \dfrac{\partial}{\partial \xi}\delta^{\Lambda}(\xi-\xi_\mu)\dfrac{\partial}{\partial \xi}\delta^{\Lambda}
(\xi-\xi_\nu)d\xi \\
&\approx& \sum_{j=0}^{N} \dfrac{\partial}{\partial\xi}\delta^{\Lambda}(\xi_j-\xi_\mu)\dfrac{\partial}{\partial\xi}\delta^{\Lambda}
(\xi_j-\xi_\nu)w^{\Lambda}_j\, . \nonumber
\end{eqnarray}
The matrix $S^{\Lambda}$ can be written as a product,
\begin{eqnarray*}
S^{\Lambda} = D(w)D^{\dagger}(w)
\end{eqnarray*}
with 
\begin{eqnarray}
\label{D}
D_{i,j}(w) = \dfrac{\partial}{\partial \xi}\delta^{\Lambda}(\xi_i-\xi_j)
\sqrt{w^{\Lambda}_i}
\end{eqnarray}
and the derivatives given in Eq.~\eqref{eq17}.
Recall that at the Gauss-Lobatto-Legendre points, the cardinal functions, cf.~Eq.~\eqref{eq16}, obey 
\begin{eqnarray*}
 \delta^k(r_i-r_j) = \delta_{i,j}\, ,
\end{eqnarray*}
where $\delta_{i,j}$ stands for the Kronecker delta. This, together with 
Eq.~\eqref{eq21}, yields the following algebraic expression 
\begin{widetext}
\begin{eqnarray}
\sum_{j=0}^{N} u^k_j\left(\frac{\hbar^2}{2\mu} \mathcal{J}^{-1}_k S^{\Lambda}_{i,j} + V(r_j)\delta_{i,j}w^k_j \right)
+\frac{\hbar^2}{2\mu}  \nabla u(r^k_0) \delta_{0,i}
-\frac{\hbar^2}{2\mu} \nabla u(r^k_N) \delta_{N,i} = \lambda\sum_{i=0}^N u^k(r_i)w^k_j\delta_{i,j}\,,\quad\quad
\label{eq35}
\end{eqnarray}
\end{widetext}
with $i=0,\ldots,N$ for the weak form of the Schr\"odinger equation, Eq.~\eqref{eq32}, within the domain $\Omega_k$.

\subsection{Global representation}
\label{subsec:global}

Finally, we need to assemble all domains $\Omega_k$, $k=1,\ldots,M$, in order to construct a global representation of the time-independent Schr\"odinger equation~\eqref{eq:tise}, and thus the Hamiltonian, from Eq.~\eqref{eq35}.
Since $\Omega = \bigcup_{k = 1}^M \Omega_k$, this can simply be done by adding the multi-domain bilinear forms defined in Eqs.~\eqref{eq:bilinear},
\begin{eqnarray}
\label{eq36}
a(u,v) = \lambda (u,v) \Leftrightarrow\sum_{k=1}^{M} a^k(u,v)
=\sum_{k=1}^{M} \lambda(u,v)_{\Omega_k}  \,,
\end{eqnarray}
provided that the correct boundary conditions are ensured at the intersection of two contiguous elements, 
\begin{subequations}\label{eq37}
\begin{alignat}{2}
 & r^k_N && \overset{!}{=} r^{k+1}_0 \label{eq37a}\,, \\
 & u^k(r)\big |_{r=r^k_N} && \overset{!}{=} u^{k+1}(r)\big |_{r=r^{k+1}_0}  \label{eq37b} \,,\\
 & \nabla u^k(r)\big|_{r=r^k_N} && \overset{!}{=}  \nabla u^{k+1}(r)\big |_{r=r^{k+1}_0} \label{eq37c}\, .
\end{alignat}
\end{subequations}
Continuity and differentiability of the global solution need to be enforced since the global cardinal basis, defined as $v^k(r-r^k_N)\cup v^{k+1}(r-r^{k+1}_0)$, is not differentiable at the $M-1$ interelement points. 
Consider the sum of $a^k(u,v)$ for two contiguous elements,
\begin{widetext}
\begin{eqnarray}
\nonumber a^k(u^{k},v^{k})  + a^{k+1}(u^{k+1},v^{k+1})
  &=& \frac{\hbar^2}{2\mu} \int_{\Omega_k} \nabla u^k\left(r\right)\nabla v^k\left(r\right)dr + b_{\Omega_k\cup\Omega_{k+1}}
+\frac{\hbar^2}{2\mu} \int_{\Omega_{k+1}} \nabla u^{k+1}\left(r\right)\nabla^{k+1}v\left(r\right)dr \\
&&+\frac{\hbar^2}{2\mu} \left( v^{k}\left(r^k_0\right)\nabla u^{k}\left(r^k_0\right)
- v^{k+1}\left(r^{k+1}_N\right)\nabla u^{k+1}\left(r^{k+1}_N\right)\right)\nonumber\\
&&+\frac{\hbar^2}{2\mu}  
\left( v^{k+1}\left(r^{k+1}_0\right) \nabla u^{k+1}\left(r^{k+1}_0\right) -
v^{k}\left(r^k_N\right) \nabla u^{k}\left(r^{k}_N\right) \right)\,,\label{eq38}
\end{eqnarray}
where we have defined
\begin{eqnarray*}
b_{\Omega_k\cup\Omega_{k+1}} =  b^k(u^k,v^k)+ b^{k+1}(u^{k+1},v^{k+1})\quad
\mbox{ with }\quad
b^k(u^k,v^k) = \int_{\Omega_k} u^k(r)V(r)v^k(r)\, dr\,.
\end{eqnarray*}
\end{widetext}
For the bilinear form $a(u,v)$, the condition of differentiability implies that the last term in Eq.~\eqref{eq38} vanishes. 
Thus, when adding the bilinear forms for all intervals $\Omega_k$, 
the interelement boundary conditions cancel out, as desired.

Analogously to Eq.~\eqref{eq37} for the bilinear forms, we introduce the global interpolant $u(r)$ as 
\begin{eqnarray}
\label{eq41}
u(r) := \sum_{k=1}^M u^k(r) = \sum_{k=1}^M \sum^{N}_{j=0}
u^k(r^k_j)v^{k}_j(r)\, .
\end{eqnarray}
Then, just as the basis set expansion of $u^k(r)$, Eq.~\eqref{eq24}, has led to $N+1$ algebraic equations within the interval $\Omega_k$, Eq.~\eqref{eq41} results in $M\times(N+1)$ algebraic equations for the total domain $\Omega$,
\begin{eqnarray}
\sum_{k=1}^M\sum_{k^\prime=1}^M\sum_{j=0}^N u^{k^\prime}_j 
a^k(v^{k^\prime}_j,v^q_i) &=&\lambda \sum_{k=1}^M\sum_{k^\prime=1}^M\sum_{j=0}^N u^{k^\prime}_j(v^{k^\prime}_j,v^q_i)_{\Omega_k}\nonumber\\
\label{eq42}
\end{eqnarray}
with $i=0,\ldots,N$, $q=1,\ldots M$. Note that the subscripts $i,j$ run over the collocation points whereas the superscripts $k,k^\prime, q$ indicate the intervals. Since the cardinal functions $\delta^k(r-r_j)$ are non-zero only within their own interval $\Omega_k$, we find
\begin{eqnarray}
\label{adelta}
a^k(v^p_j,v^q_i) &=& a^k(v^k_j,v^k_i)\delta_{k,p}\delta_{p,q}
\end{eqnarray}
and
\begin{eqnarray}
\label{mdelta}
(v^p_j,v^q_i)_{\Omega_k} &=& w^k_i\delta_{k,p}\delta_{p,q}\delta_{i,j}\,.
\end{eqnarray}
Therefore, Eq.~\eqref{eq42}  takes the same form as Eq.~\eqref{eq35} but with $(N+1)\times(M-1)$ vanishing terms. In other words, the global representation, by construction, is equivalent to writing the elemental equation~\eqref{eq35} $M\times(N+1)$ times, i.e., as many times as there are  
configurations for the test function $v^k_j(r)$ with $j=0,\ldots,N$ and  $k=1,\ldots,M$, while accounting for the boundary conditions~\eqref{eq37}.
Specifically, when adding the two algebraic equations for $q=k,j=N$ and $q=k+1, j=0$, for $k=1,M-1$, in Eq.~\eqref{eq42}, the last (vanishing) term in Eq.~\eqref{eq38} is retrieved at the $M-1$ connection points. We thus obtain a system of $M\times(N+1) - (M-1) = N\times M+1$ algebraic equations, in accordance with the number of collocation points in the global representation, i.e., without any repetition of points.  

Solving the linear system of equations~\eqref{eq42} with the boundary conditions~\eqref{eq37}
is then equivalent to solving the generalized eigenvalue problem
\begin{eqnarray}
\label{mass}
A\, u = \lambda\, \mathcal{M}(w)\, u\,, 
\end{eqnarray}
where $\mathcal{M}(w)$ is a $(N\times M+1)\times (N\times M+1)$ diagonal
matrix, hereafter
referred to as the \textit{global mass matrix}. Its matrix elements are
given in terms of the Gaussian quadrature weights $w^k_j$, cf. Eq.~\eqref{eq9},
\begin{subequations}
\begin{eqnarray}
\label{Melements}
\mathcal{M}_{i,i}(w) = \gamma^k_j(w),\hspace{0.2cm} i = N(k-1) + j +1\, ,
\end{eqnarray}
with $j=0,\ldots,N$, $k=1,\ldots,M$, and 
\begin{eqnarray}
\label{eq44}
\gamma^k_j(w) =
\begin{cases}
w^k_j& \mbox{if }  k < M   \mbox{ and }  0 < j < N\,, \\
w^k_N+w^{k+1}_0 & \mbox{if }  k \le M  \mbox{ and }  j = 0\,, \\
w^{k-1}_N+w^{k}_0 & \mbox{if } k < M   \mbox{ and } j = N\,, \\
w^1_0 & \mbox{if } k = 1   \mbox{ and }  j = 0\,, \\
w^M_N & \mbox{if } k = M   \mbox{ and }  j = N\, . \\
\end{cases}
\end{eqnarray}
\end{subequations}
Note that the weights defined at the interelement points, i.e., $x^k_{N}$ and  $x^{k+1}_{0}$, are defined as $ w^k_N+w^{k+1}_0$. This can be easily shown by using the additivity theorem of integration for continuous functions.
The matrix $A$ corresponds to the global  representation of the bilinear form
$a(u,v)$. Because of the compact support of the basis functions $v_j^k(r)$, 
$A$ is characterized by a sparse structure, with matrix elements 
\begin{subequations}
\begin{eqnarray}
\label{A}
A_{i,j} &=&
\begin{cases}
a^k(v^k_{i^\prime},v^k_{j^\prime}) & \mbox{if } i^\prime\ne j^\prime\ne 0 \mbox{ or }  i^\prime\ne j^\prime\ne N\,, \\
a^1(v^1_0,v^1_0) & \mbox{if }  k = 1 \,, \\
a^M(v^M_N,v^M_N) & \mbox{if }  k = M \,,\\
a^{k,k+1} & \mbox{if }  k< M  \mbox{ and } i^\prime= j^\prime=N\,, \\
a^{k-1,k} & \mbox{if }  k\ge 2 \mbox{ and }i^\prime= j^\prime= 0 \,,\\
0 & \mbox{otherwise}\,,
\end{cases}\quad
\end{eqnarray}
and global indices 
\begin{equation}
  \label{eq:Aii}
  i=N(k-1)+i^\prime+1  \mbox{ and } j=N(k-1)+j^\prime+1  \,,
\end{equation}
 such that  $1\le i,j\le NM+1$ for $i^\prime,j^\prime = 0,\ldots,N$ and 
\begin{eqnarray}
\label{a_interface}
a^{k,k+1} = a^k(v^k_N,v^k_N)+a^{k+1}(v^{k+1}_0,v^{k+1}_0)\, .
\end{eqnarray}
The \textit{elemental} bilinear form $a^k(v^k_{i^\prime},v^k_{j^\prime})$,\ is given by 
\begin{eqnarray}
\label{aj2a}
a^k(v^k_{i^\prime},v^k_{j^\prime}) &=& \frac{\hbar^2}{2\mu}\mathcal{J}^{-1}_k S^{\Lambda}_{i^\prime,j^\prime} + V(r_{i^\prime})\delta_{i^\prime,{j^\prime}}w^k_{j^\prime} \\
&&+\frac{\hbar^2}{2\mu} \Big( \nabla u(r^k_0) \delta_{0,{j^\prime}} 
-  \nabla u(r^k_N) \delta_{N,{j^\prime}}\Big)\, ,\nonumber
\end{eqnarray}
\end{subequations}
where  $S^{\Lambda}_{i,j}$ is defined in Eq.~\eqref{eq21} and $\mathcal{J}^{-1}_k$ refers to the inverse of the Jacobian~\eqref{eq:Jac}.

Solution of Eq.~\eqref{mass} requires significantly less numerical effort, if  $\mathcal{M}$ can be transformed into identity. To this end, it suffices to renormalize the basis functions,
\begin{eqnarray}
\label{PN2}
\tilde v^k_j(r)=\frac{\delta^{k}(r-r_j)}{\sqrt{\gamma^k_j}},\, 
\end{eqnarray}
Then, Eq.~\eqref{eq24}, i.e., the  solution of Eq.~\eqref{eq29}, takes the following form 
\begin{subequations}
\begin{eqnarray}
\label{e43}
u^{k}(r) = \sum_{j=0}^{N}\tilde{u_j}^{k}(r)\tilde{v}^k_j(x)\, , 
\end{eqnarray}
with 
\begin{eqnarray}
\label{utilde}
\tilde{u}^k_j := u^k_j\times\sqrt{\gamma^k_j}\,. 
\end{eqnarray}
\end{subequations}
Using Eq.~\eqref{e43}, the linear system of equations~\eqref{eq42} 
becomes
\begin{eqnarray}
\label{eq42tilde}
\sum_{k=1}^M\sum_{k^\prime=1}^M\sum_{j=0}^N \tilde{u}^{k^\prime}_j a^k(\tilde{v}^{k^\prime}_j,\tilde{v}^q_i) =
\lambda \sum_{k=1}^M\sum_{{k^\prime}=1}^M\sum_{j=0}^N \tilde{u}^{k^\prime}_j(\tilde{v}^{k^\prime}_j,\tilde{v}^q_i)_{\Omega_k}\quad\quad
\end{eqnarray}
which is equivalent to solving 
\begin{eqnarray}
\label{masstilde}
\tilde{A}\, \tilde{u} = \lambda\, \tilde{u} 
\end{eqnarray}
with matrix elements 
\begin{eqnarray}
\label{Hamiltonian}
\tilde{A}_{i,j} = 
\dfrac{A_{i,j}}{\sqrt{\gamma^k_{i^\prime}\, \gamma^k_{j^\prime}}}\, ,
\end{eqnarray}
where $A_{i,j}$ is given in Eq.~\eqref{A}.
The actual value of the eigenfunction at  $r=r^k_j$ is obtained as $u^k_j = \tilde{u}^k_j/\sqrt{\gamma^k_j}$.

In order to explicitly state the global boundary conditions, it is convenient to rewrite Eq.~\eqref{masstilde} in the following form,
\begin{subequations}
\begin{eqnarray}
\label{eq:H0}
\tilde{H}\, \tilde{u} = \lambda\,\tilde{u} + \tilde{A}^{(0)}\, \tilde{u} \,,
\end{eqnarray}
where $\tilde{A}=\tilde{H}-\tilde{A}^{(0)}$ and  $\tilde{A}^{(0)}\tilde{u}$ denoting the boundary condition vector, 
\begin{eqnarray}
\label{u0}
  \Big(\tilde{A}^{(0)}\tilde{u}\Big)_{i}
  &=& \frac{\hbar^2}{2\mu}\Big(- \nabla u(r^k_{i^\prime})\delta_{0,i^\prime}\delta_{1,k} + \nabla u(r^k_N)\delta_{N,i^\prime}\delta_{M,k} \Big)\nonumber \\
  &=& \frac{\hbar^2}{2\mu}\Big( \nabla u(r_1)\delta_{i,1} - \nabla u(r_{_ {NM+1}}) \delta_ {_{i,NM+1}}\Big)\quad
\end{eqnarray}
with $i=i(i^\prime,k)$  found in Eq.~\eqref{eq:Aii}.
In particular for bound states and eigenstates in a box, it is required that
\begin{eqnarray*}
u^1_0 = u^M_N = 0\,.
\end{eqnarray*}
\end{subequations}
This can be enforced by the choice of basis functions, i.e., by 
ensuring $v^1_0 \overset{!}{=}0$ and $v^M_N \overset{!}{=} 0$.
A simple implementation is achieved by taking $j=1,\dots, N$ for $k=1$ and $j=0,\dots,N-1$ for $k=M$ instead of $j=0,\dots, N$. The matrix representation of the Hamiltonian is then given by 
\[\tilde{H}_{i-1,j-1}=\tilde{A}_{i,j}\,, \quad i,j=2,\ldots, NM\,.
\] 
For Dirichlet boundary conditions, Eq.~\eqref{eq:H0} takes thus the form 
\begin{eqnarray*}
\tilde{H}\, \tilde{u} &=& \lambda\, \tilde{u}\,.
\end{eqnarray*}
Despite the dense structure of the matrix representation of the kinetic operator in each interval $\Omega_k$, the local support of the basis functions $v^k_j(r)$ translates into a global kinetic energy matrix that is blockwise sparse except for the interelement points, 
\begin{widetext}
\begin{eqnarray}
\label{sparse}
H = 
\left(
  \begin{array}{ccccccccccccc}
    \cdots &\tikzmarkin[line width=0.4pt]{a} a^{k-2}_{N,N}+a^{k-1}_{0,0} &a^{k-1}_{0,1} &\cdots &a^{k-1}_{0,N} &0 &0 &0 &0 & \cdots&0&0&\cdots \\
\cdots & a^{k-1}_{0,1} & a^{k-1}_{1,1} & \cdots & a^{k-1}_{1,N}& 0 &0  &0  &0 &\cdots & 0&0&\cdots\\
\cdots & \vdots &\vdots & \ddots&\vdots &0 &0 &0 &0 &\cdots &0&0&\cdots \\
    \cdots &a^{k-1}_{N,0} &a^{k-1}_{N,1} &\cdots & \tikzmarkin[line width=0.4pt]{b}\tikzmarkin{e} a^{k-1}_{N,N}+a^{k}_{0,0}\tikzmarkend{e} \tikzmarkend{a} 
  &\textcolor{black}{a^{k}_{0,1}} &\cdots &\textcolor{black}{a^{k}_{0,N}} &0 & \cdots &0&0&\cdots \\
\cdots &0 &0 &0&\textcolor{black}{a^{k}_{1,0}} &\textcolor{black}{a^{k}_{1,1}} &\cdots &\textcolor{black}{a^{k}_{1,N}} &0 &\cdots &0&0&\cdots \\
\cdots &0 &0 &0 &\textcolor{black}{\vdots} &\textcolor{black}{\vdots} &\ddots &\vdots &0 &\cdots &0&0&\cdots \\
    \cdots&0 &0&0 &\textcolor{black}{a^{k}_{N,0}} &\textcolor{black}{a^{k}_{N,1}} & \cdots &\tikzmarkin[line width=0.4pt]{c} \tikzmarkin{f} a^{k}_{N,N}+a^{k+1}_{0,0} \tikzmarkend{f} \tikzmarkend{b} & \textcolor{black}{a^{k+1}_{0,1}} &\cdots &\textcolor{black}{a^{k+1}_{0,N}}&0&\cdots \\
 \cdots&0 &0 &0 &0&0 &\cdots &\textcolor{black}{a^{k+1}_{1,0}} &\textcolor{black}{a^{k+1}_{1,1}} &\cdots &\textcolor{black}{a^{k+1}_{1,N}}&0&\cdots \\
 \cdots&0 &0 &0 &0 &\vdots &\vdots &\textcolor{black}{\vdots} &\textcolor{black}{\vdots} &\textcolor{black}{\ddots}&\vdots&0&\cdots \\
    \cdots&0 &0 &0 &0&0&0 &\textcolor{black}{a^{k+1}_{N,0}} &\textcolor{black}{a^{k+1}_{N,1}} &\textcolor{black}{\cdots} &              a^{k+1}_{N,N}+a^{k+2}_{0,0} \tikzmarkend{c}  &\textcolor{black}{a^{k+2}_{0,1}}&\textcolor{black}{\cdots} \\ 0&0&0&0&0&0&0&0&0&0&\textcolor{black}{a^{k+2}_{1,0}}&a^{k+2}_{1,1}&\cdots\\ 
 0&0&0&0&0&0&0&0&0&0&\vdots&\vdots&\ddots\\
 \end{array}
 \right)
 \end{eqnarray}
\end{widetext}

The sparsity and band-like structure can be exploited to reduce storage and CPU time in both diagonalization and time propagation, using standard libraries for sparse matrix-vector operations~\cite{Lehoucq97arpackusers}. The number of matrix elements that need to be stored when exploiting the band-like structure 
is found to be 
\begin{eqnarray}
  \mathcal{N}_{spar} &=& (N\times M+1)(N+1) - N\left(N+1\right)/2 - 2(N+1)\nonumber\\
 \label{eq:storage_spar}
\end{eqnarray}
where $N+1$ is the number of collocation points and $M$ denotes the number of intervals. This compares to the $N_{pts}\left(N_{pts}+1\right)/2$ different matrix elements of a full Hermitian matrix.

\subsection{Choice of sub-intervals}
\label{subsec:split}
The remaining question is how to choose the sub-intervals $\Omega_k=[r_0^k,r_N^k]$. As stated in Section~\ref{subsec:weak}, there is no a priori restriction on $r^k_0$ and $r^k_N$ for all $k = 1,\ldots, M$. Here, we utilize the intuition underlying the Mapped Fourier Grid method~\cite{KokoulineJCP99,FattalPRE96,KallushCPL06,TiesingaPRA98} and adapt the size of $\Omega_k$ to the local de Broglie wavelength. This implies that $\Omega_k$ gets larger in the asymptotic part of the potential. 

It is achieved as follows. The overall domain $\Omega$ starts at $r_{min}$, i.e.,  $r^k_0=r_{min}$ for $k=1$. The upper edge of the first interval, $r^{k=1}_N$, is obtained by solving the implicit equation~\cite{KallushCPL06}
\begin{equation}
  \label{eq:MFGH}
  \beta = \frac{\sqrt{2\mu}}{\pi} \int^{r^k_N}_{r^k_0} \sqrt{E_{asy}-V(r)} dr\,,
\end{equation}
where $\beta$ and $E_{asy}$ are two prespecified constants. For all further intervals $\Omega_k$, $r^{k+1}_0$ is set equal to $r^{k}_N$, and  $r^{k+1}_N$ is obtained by solving Eq.~\eqref{eq:MFGH}. This procedure is repeated until $r_{max}$ is reached.   

The two constants in  Eq.~\eqref{eq:MFGH} have a physical meaning, making their choice straightforward. The parameter $\beta$, $0 < \beta \le 1$, estimates the local coverage of the phase space volume~\cite{KallushCPL06}: Smaller values of $\beta$ result in a higher density of points, and $\beta = 1$ corresponds to the minimal classical estimation for the phase space discretization. The parameter $E_{asy}$ specifies the smallest energy up to which the size of $\Omega_k$ is increased---if the grid includes $r$ for which $V(r)$ is smaller than $E_{asy}$, the size of the intervals is kept constant. 

\begin{figure}[tb]  
\centering
\vspace{-0.1cm}
\includegraphics[width=1.00\linewidth]{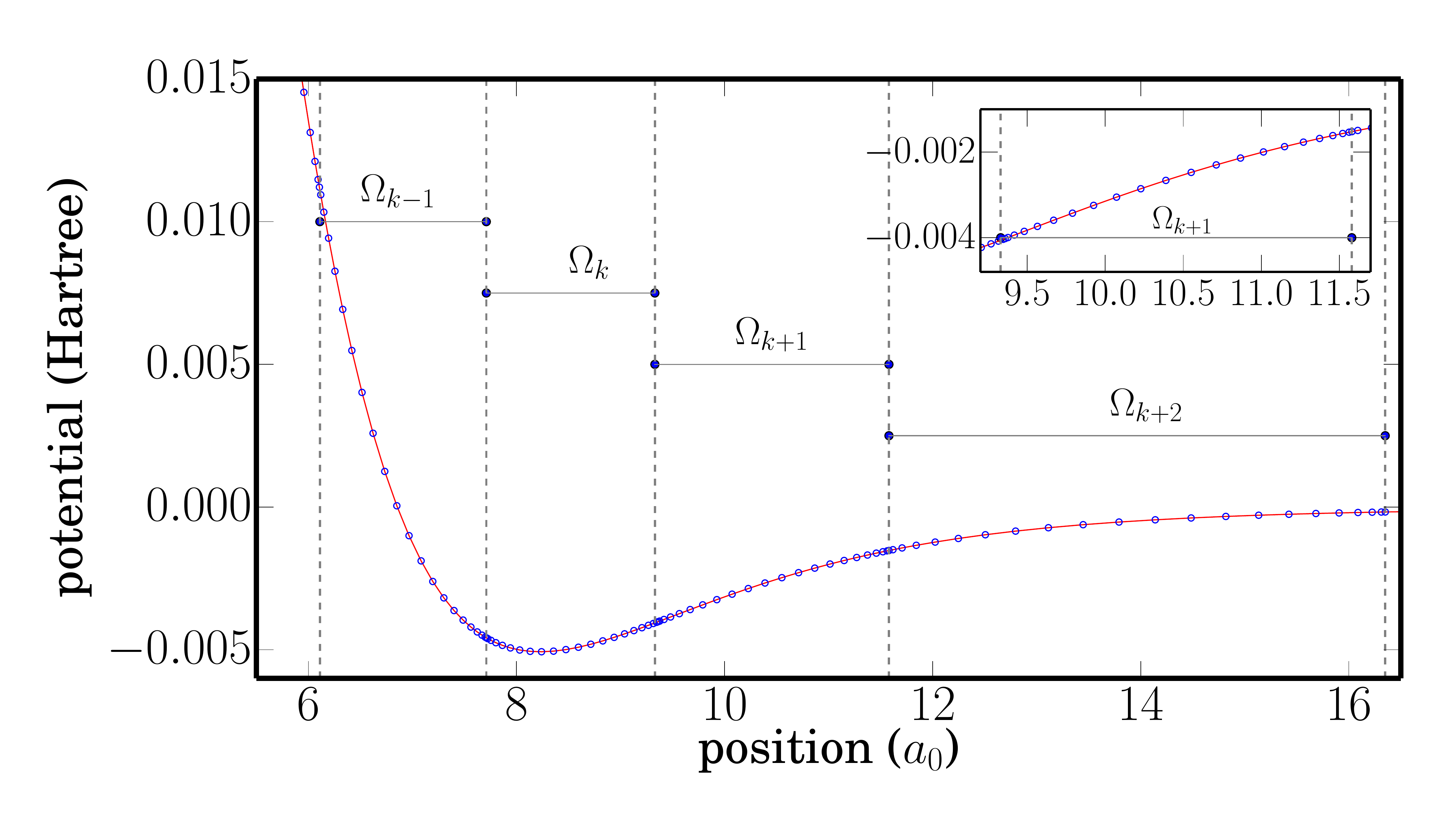}
\caption{Distribution of collocation points with $N=21$. The inset shows a zoom onto the interval labeled by $\Omega_{k+1}$. The high density of points close to the edges of the interval is typical for collocation based on Gauss-Lobatto-Legendre points.}
\label{fig:MFHG_SEM}
\end{figure}
Within each interval $\Omega_k$, the points $r^k_j$, $j=0,\ldots,N$, are chosen according to the Legendre quadrature rule, as described in Section~\ref{subsec:colloc}. Since each interval $\Omega_k$ is discretized by  $N+1$ collocation points, the density of points per element is constant. 
The resulting discretization is illustrated in Fig.~\ref{fig:MFHG_SEM} for the B$^{1}\Sigma^+_u$ electronically excited state of the Ca$_2$ molecule~\cite{KochPRA08} which vanishes asymptotically as $1/R^3$.  Such long-range states support extremely weakly bound vibrationally levels and therefore require large $r_{max}$ to faithfully represent all bound levels~\cite{KokoulineJCP99}. Such levels are relevant for example in the photoassociation of ultracold atoms, and it was the need to calculate such levels that had prompted the development of the mapped Fourier grid method~\cite{KokoulineJCP99}. 
We will analyze the accuracy as well as the computational resources for the calculation of such levels with our multi-domain pseudospectral approach and compare it to the mapped Fourier grid method in Section~\ref{sec:accuracy} below before applying it in time-dependent calculations in Section~\ref{sec:HHG}.

\section{Choice of domain number and collocation order}
\label{sec:accuracy}

The two parameters which are crucial for the analysis of accuracy and efficiency of the multi-domain pseudospectral approach 
are the number of intervals, $M$, and the number of collocation points within each interval $N+1$, or equivalently, the order of the interpolation polynomial, $N$. 
If $M$ and $N$ are chosen optimally, the calculation will be highly accurate while minimizing at the same the requirements on storage and CPU time.
The role of $M$ and $N$ in our approach is similar to the parameters $h$ and $p$ in  finite element methods~\cite{boyd,GuoCompMech86,Babuvska1982,RankNME2001}, where  the standard $h$-version, also referred to as $h$-refinement~\cite{boyd},
consists in keeping the degree of the interpolating polynomials, usually of low 
degree, $p=1,2$, unchanged while modifying the size of each 
subdomain~\cite{boyd,GuoCompMech86,Babuvska1982,RankNME2001}. Alternatively, the $p$-version, consists in keeping the size of each element unchanged while increasing the order of the interpolating polynomials~\cite{boyd,GuoCompMech86,Babuvska1982,RankNME2001}. 
Finally, the $h$-$p$-version of the finite element method~\cite{GuoCompMech86} modifies the size of each
element only in regions where high resolution is needed~\cite{boyd}.

\begin{figure}[tb]
\centering
\includegraphics[width=1.05\linewidth]{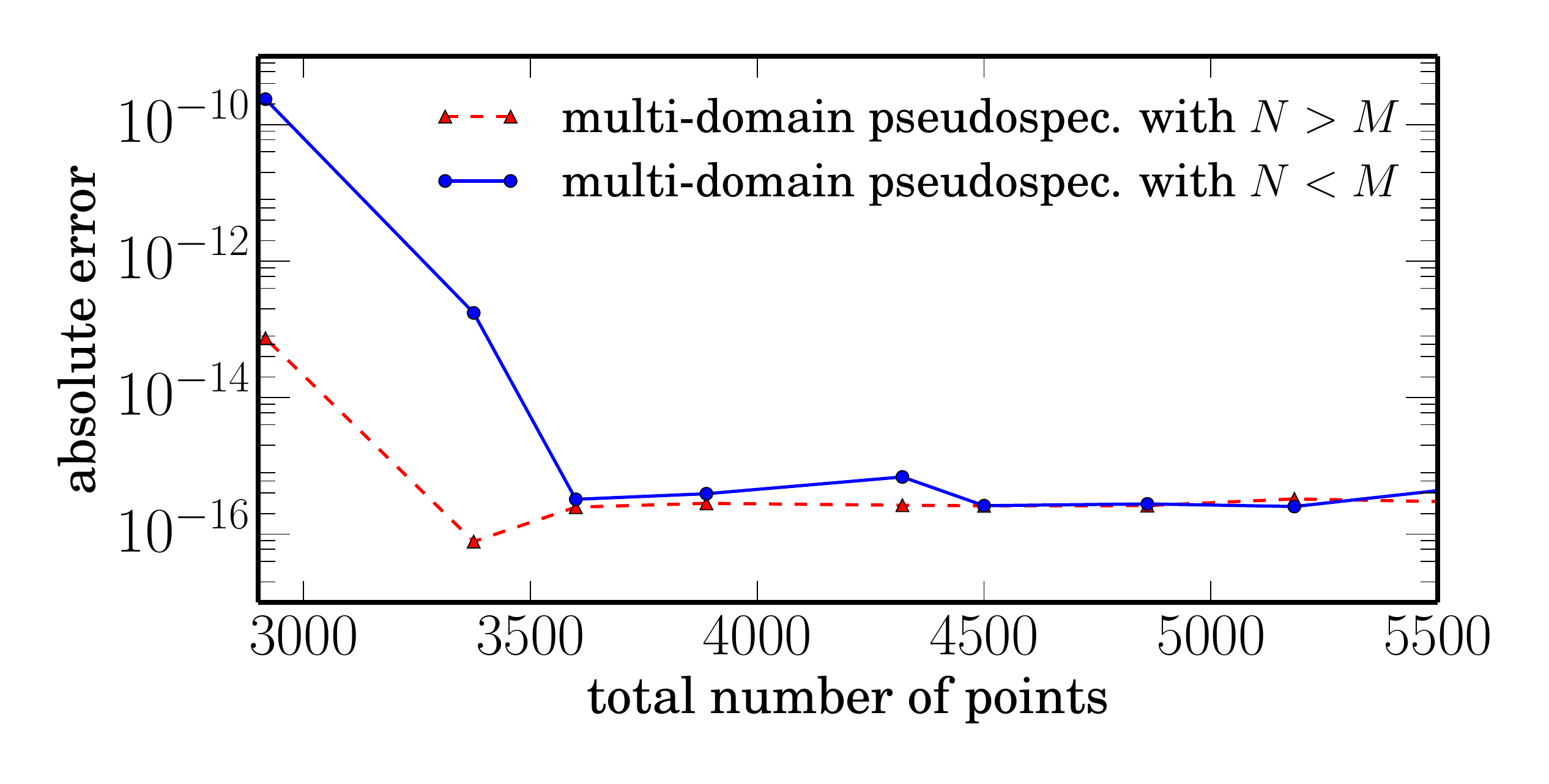}
\caption{Accuracy of the eigenvalue of a weakly bound level, calculated with 
  the multi-domain pseudospectral method, referenced to the result
  obtained with the mapped Fourier grid Hamiltonian using a large
  number of points ($N_{pts}=20000$). The reference eigenvalue is
  $E_{ref} =-2.2640245\times 10^{-10}\,$Hartree,
  compared to $E_{0} \approx -2.607\times
  10^{-2}\,$Hartree which is the eigenvalue with largest magnitude. 
  When the total number of collocation points is sufficiently large, the accuracy of the multi-domain pseudospectral method is 
  independent of the choice of the number of domains $M$ and the collocation order $N$.
}
\label{fig:nl_vs_m}  
\end{figure}
As a first practical example, we consider the calculation of a weakly bound level of the Ca$_2$ B$^{1}\Sigma^+_u$ electronically excited state.
The overall spatial domain is chosen with $r_{min}=4.5\,$a$_0$,
$r_{max}=50000\,$a$_0$. For the mapped Fourier grid Hamiltonian, we
take the total number of grid points to be $N_{pts}=20000$  which
corresponds to $\beta=0.029$. Choosing the eigenvalue labeled by
$v_{ref}=229$ with $E_{ref} = -2.2640249\times 10^{-10}\,$Hartree,  we
treat the result obtained with the mapped Fourier grid Hamiltonian and
this very large number of points as a reference to benchmark the
accuracy of the multi-domain spectral method for increasing the total
number of collocation points, see Fig.~\ref{fig:nl_vs_m}. We find the
calculation using the multi-domain spectral method to be converged to
machine precision (with an arbitrary choice of $N$ and $M$) if the total number of points, $N\times M+1$, exceeds 3000. The overall precision in Fig.~\ref{fig:nl_vs_m} is
  determined by the eigenvalue with the largest magnitude, which is
  the ground state of the Hamiltonian, with magnitude of the order of
  $10^{-2}\,$Hartree. Machine precision relative to this value amounts
  to $10^{-17}\,$Hartree. The accuracy of the pseudospectral method
  saturates somewhere about $10^{-16}\,$Hartree. The missing digit is
  most likely due to different numerical routines for diagonalization
  in the  multi-domain spectral method (with a sparse Hamiltonian matrix) and
  the mapped Fourier grid method (with a fully occupied Hamiltonian
  matrix).

An important question concerns the best choice of the parameters $M$ and $N$. 
The same total number of points, $N\times M+1$, can be realized by two different choices of $M$ and $N$. Accuracy, storage requirement and spectral radius are, however, not the same between one configuration and the other. It is known from finite-element methods, that the $p$-refinement shows better convergence than the $h$-version~\cite{boyd}. In particular when just a small number of points is used, 
the accuracy may be improved by choosing $N>M$~\cite{boyd}.  Nevertheless, 
the imbalance between $N>M$ and $N<M$ is removed when the overall number of points becomes sufficiently large, as shown in Fig.~\ref{fig:nl_vs_m}.
Remarkably, the accuracy reaches a stationary value and remains independent
of the choice of $M$ and $N$. 
The corresponding flexibility in the choice of $N$ and $M$ is crucial for choosing optimal values for time propagation. On one hand,
choosing larger $N$, i.e., a higher degree of the interpolation polynomial, 
and smaller $M$ considerably reduces the total number of grid points,
$N_{pts}$, for a given accuracy. Smaller $N_{pts}$ decreases the spectral radius. On the other hand, our numerical tests 
show that the decrease of the spectral radius is actually even faster
for the case of larger $M$ and smaller $N$ (with a correspondingly
larger  total number of points $N_{pts}$). We therefore focus on this second option and see in what follows that choosing a 
larger total number of points $N_{pts}$ (with smaller $N$  and  larger $M$) does not compromise the efficiency of the Chebychev propagation nor increase the storage requirements for the Hamiltonian matrix.

\begin{figure}[tb]
\centering
\includegraphics[width=1.05\linewidth]{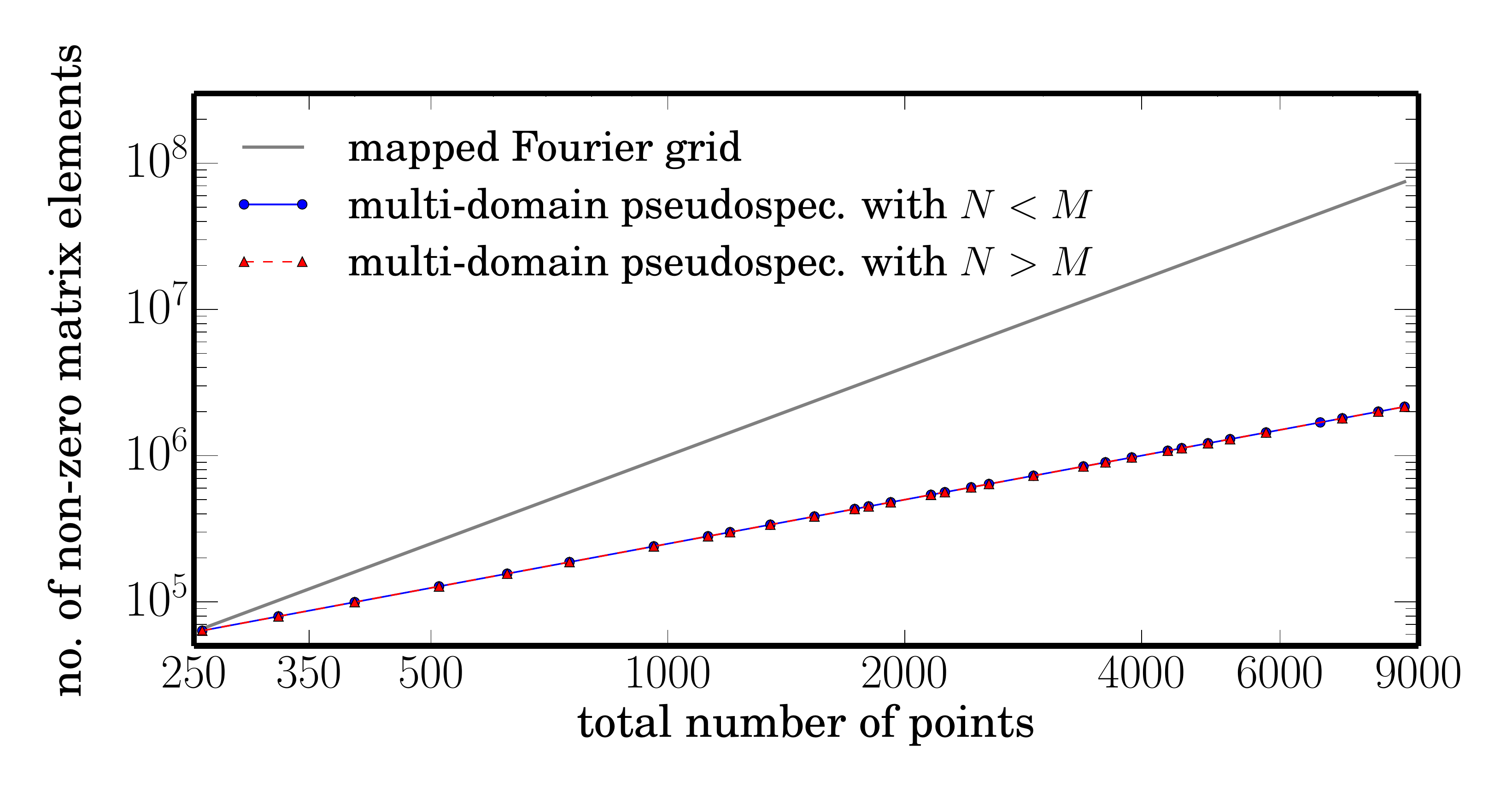}
\caption{Number of non-zero matrix elements of the Hamiltonian that need to be stored in memory. The mapped Fourier grid leads to a full kinetic energy matrix, whereas the Hamiltonian is sparse in the multi-domain pseudospectral representation. Note the log-log scale. 
}
\label{fig:plot_accuracy1}  
\end{figure}
The corresponding number of non-zero matrix elements of the Hamiltonian, i.e., the storage requirement, is shown in Fig.~\ref{fig:plot_accuracy1} as a function of the total number of points. Again, $N$ and $M$ have been chosen arbitrarily.
Due to the sparsity of the Hamiltonian, the multi-domain pseudospectral representation requires significantly less storage than the mapped Fourier grid Hamiltonian. Given the fact, that the accuracy of both methods is the same for $N_{pts}>3000$, the multi-domain pseudospectral representation allows for a dramatic reduction in the memory required to calculate the spectrum without compromising accuracy. This opens new perspectives for obtaining highly accurate weakly bound states as well as scattering states for long-range potentials, for example in coupled channels calculations, where the memory required for storing the mapped Fourier Hamiltonian quickly becomes a limiting issue~\cite{GonzalezPRA12,CrubellierNJP15}.

\begin{figure}[tb]
\centering
\includegraphics[width=1.05\linewidth]{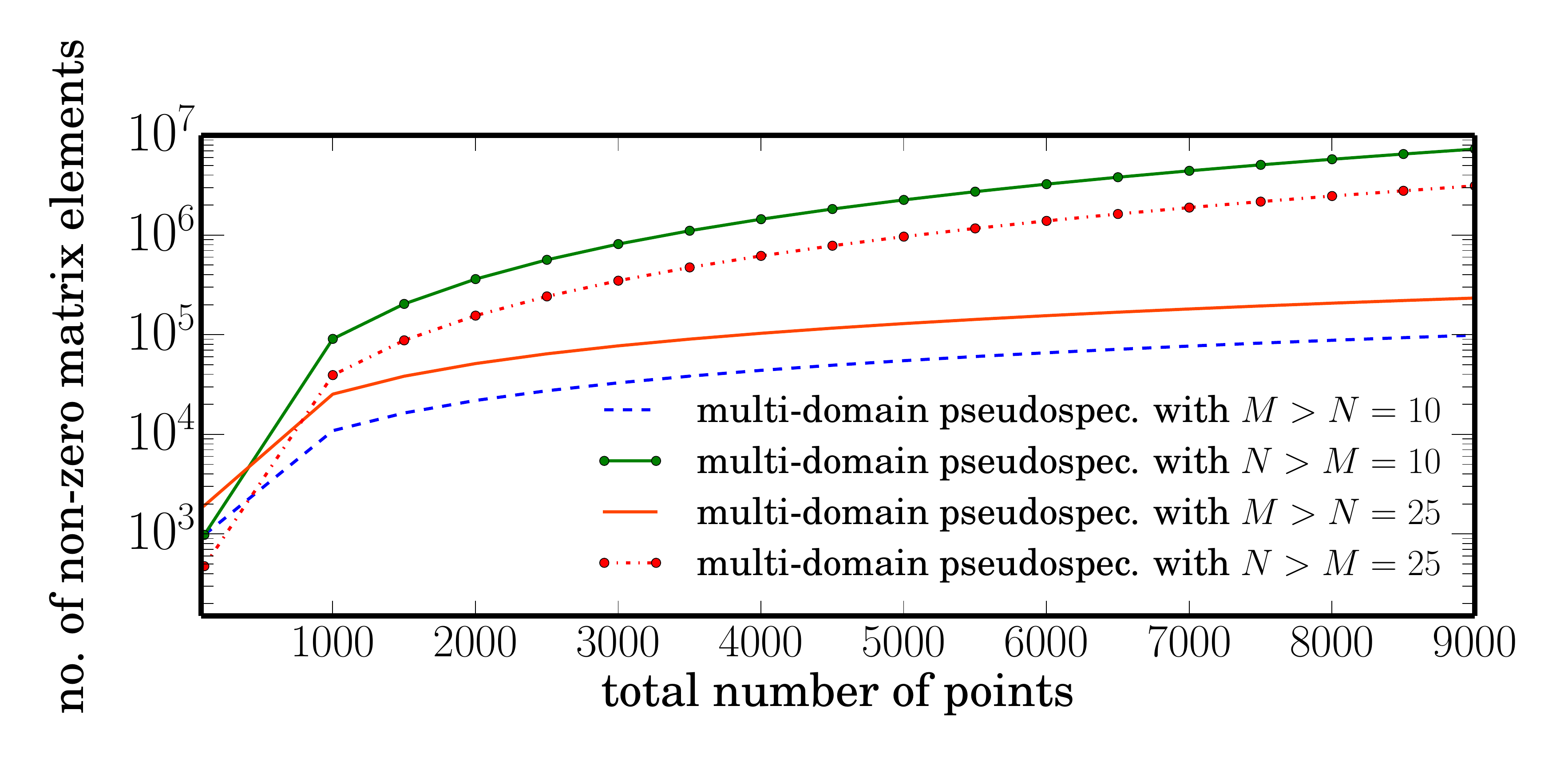}
\caption{Number of non-zero matrix elements of the Hamiltonian that need to be stored in memory for specific choices of $N$  and $M$.
}
\label{fig:figure3.5}  
\end{figure}
While different choices of $N$ and $M$ correspond to
different storage requirements, this does not show up on the scale
of Fig.~\ref{fig:plot_accuracy1}. The sparsity of the Hamiltonian is
therefore further analyzed in Fig.~\ref{fig:figure3.5} 
by comparing the cases $N>M$ and $N<M$ for a fixed number of points allowing,
this time, $N$ and $M$ to be significantly different. As  can 
be seen from Eq.~\eqref{eq:storage_spar}, for a fixed number of points
$N\times M+1$, the case $N>M$ leads to a less sparse representation of the Hamiltonian matrix.
However, both cases, $N>M$ and $N<M$, lead to a significant
improvement in terms of storage, requiring only a few percent of the
memory needed for the full matrix obtained with the mapped Fourier
grid method.

\begin{figure}[tb]
\centering
\includegraphics[width=1.05\linewidth]{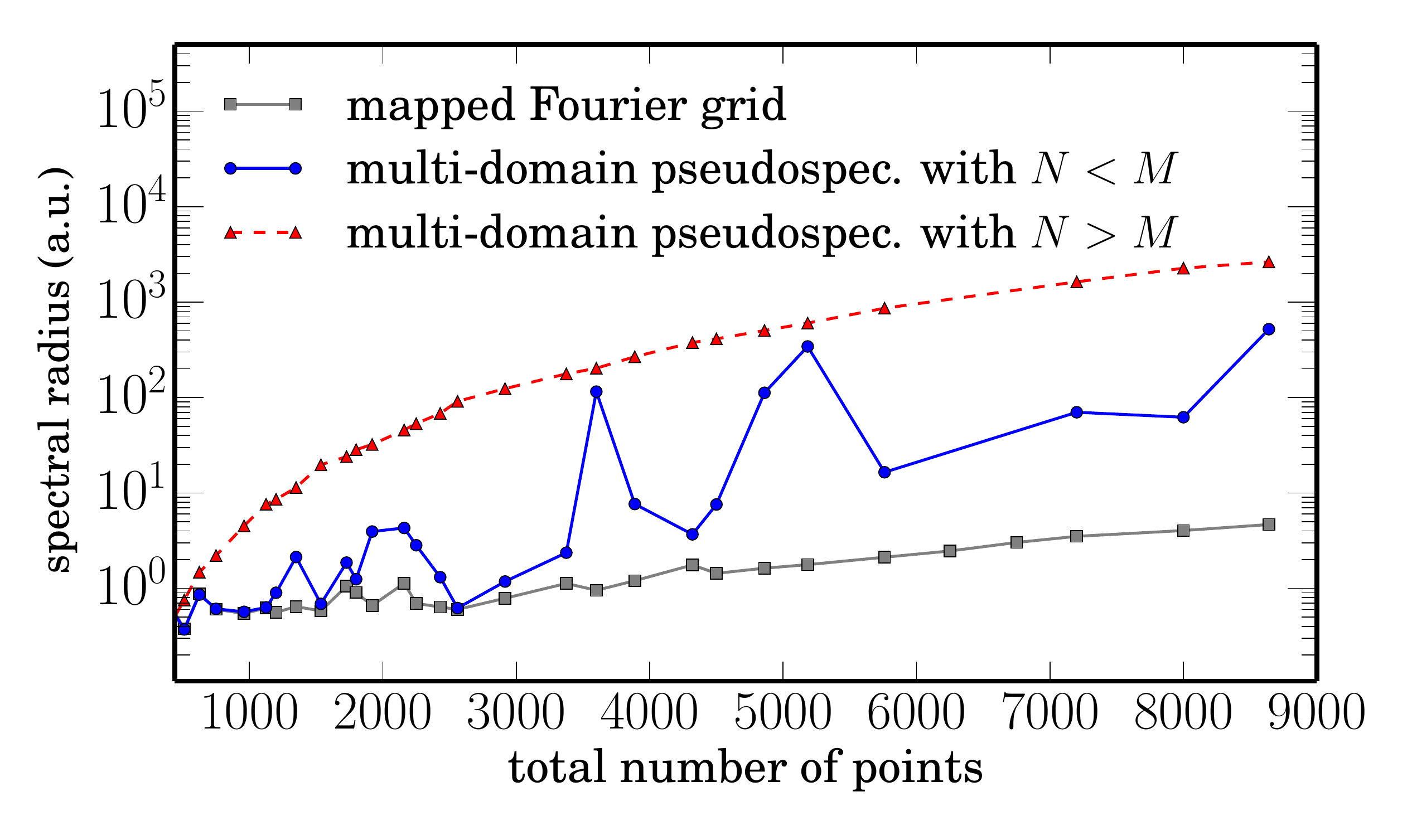}
\caption{Spectral radius as a function of the total number of points. The spectral radius determines the number of the times the Hamiltonian needs to be applied for time evolution with the Chebychev propagator.}
\label{fig:specrad}  
\end{figure}
Finally, we compare the spectral radius, $\Delta E$, obtained with the mapped Fourier grid Hamiltonian and the adaptive multi-domain pseudospectral approach in Fig.~\ref{fig:specrad}. This is important because the spectral radius determines the number of terms in the Chebychev expansion of the time evolution operator, cf. Section~\ref{sec:method}, i.e., the number of times the Hamiltonian is applied to a wavefunction. As a rule of thumb,  the spectral radius of the mapped Fourier grid Hamiltonian is smaller than that obtained with the adaptive multi-domain pseudospectral approach for the 
same number of points. Moreover, we find that for the same $N_{pts}$, the spectral radius for $N>M$ is larger than that for  $N<M$. This is
somewhat unfortunate since for a given total number of points better accuracy is obtained with $N>M$. However, since, for sufficiently large $N_{pts}$, 
the accuracy is independent of the choice of $N$ and $M$, cf. Fig.~\ref{fig:nl_vs_m}, and time propagation will be most efficient for $\Delta E$ as small as possible, it is convenient to choose a relatively large total number of points with a low order $N$ of the interpolation polynomial. This allows to reduce the numerical effort of the multi-domain pseudospectral method compared to the mapped Fourier grid Hamiltonian while keeping the level of accuracy, even though the total number of grid points required for the multi-domain pseudospectral approach  is larger than that required for the mapped Fourier grid. 

To summarize, it is optimal to (i) choose a low order of the interpolation polynomial or, equivalently, number of collocation points per element, e.g. $N = 3,4,5$, since it results in a small spectral radius, (ii) increase the total number of points such that the desired accuracy is obtained and (iii) define the number of intervals $M$ according to $N_{pts} = N\times M +1$.  

\begin{figure}[tb]
\centering
\includegraphics[width=1.00\linewidth]{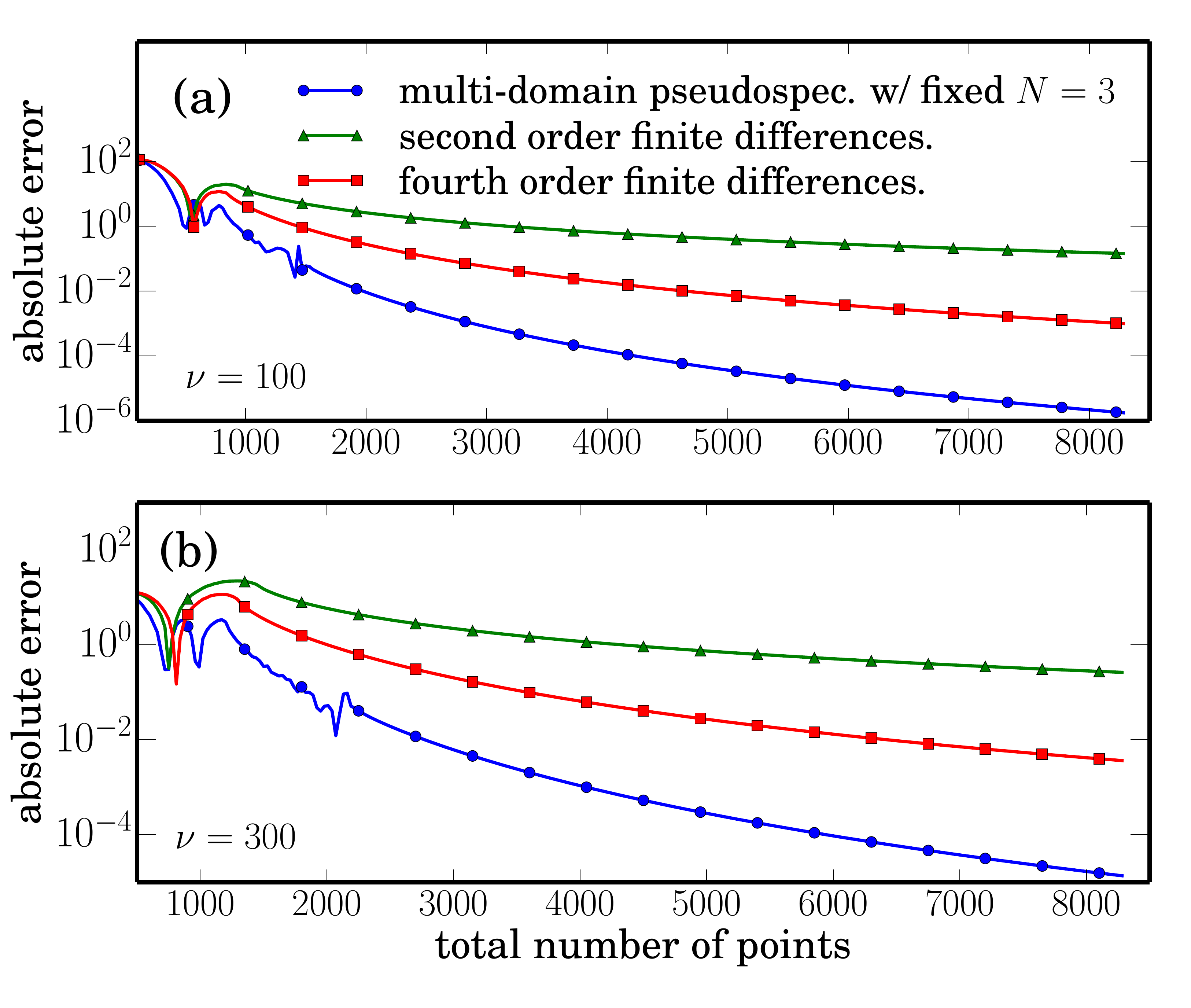}
\caption{Accuracy of the multi-domain pseudospectral approach for a low collocation order ($N=3$) compared to second and fourth orders finite differences for the levels $\nu=100$ (a) and $\nu=300$ (b) of the Morse 
  potential with eigenvalues 
  $E_{100}=-112.1253125\,$a.u. and $E_{300}=-12.3753125\,$a.u., respectively. Despite the similar structure of the Hamiltonian matrix, the pseudospectral approach is significantly more accurate. 
}
\label{fig:compare_vs_sods}  
\end{figure}

Note that for a low order of the interpolation polynomials, e.g. $N = 3$, the sparse band-like structure of the kinetic energy matrix is quite similar to what is obtained using second and fourth order finite differences. 
We therefore compare the accuracy obtained with the multi-domain pseudospectral approach for low order of the interpolation polynomials to that of the second and fourth order finite differences. As shown in Fig.~\ref{fig:compare_vs_sods}, the multi-domain pseudospectral representation yields a significantly better accuracy than  finite differences. This reflects the \textit{global} approximation of the derivatives within each interval and emphasizes the superiority of pseudo-spectral approaches over methods based on the  Taylor expansion.
\section{Application to  high harmonic generation}
\label{sec:HHG}

We now apply our adaptive-size multi-domain pseudospectral propagation method 
to simulate  high order harmonic generation. To this end, we consider
an electron subject to a soft Coulomb potential~\cite{JoachainBook2012},
\begin{eqnarray}
\label{eq:potential}
  V(x) = -\dfrac{1}{\sqrt{a+x^2}}\,.
\end{eqnarray}
The electron is subject to a linearly polarized electric field of the form  
\begin{eqnarray}
E(t)=E_0\,G(t)\sin(\omega_0\,t)\,, 
\label{eq:electric_field}
\end{eqnarray}
where $G(t)$ is a Gaussian envelope of full width at half maximum
$\tau_{FWMH}=206.5\,$a.u., the maximal field amplitude is
$E_0=0.06\,$a.u., and the carrier frequency $\omega_0 = 0.1\,$a.u. The
interaction of the electron with the electric field is treated in the
dipole approximation, 
\begin{eqnarray}
  H_I(x,t) = -xE(t)\,. 
\end{eqnarray}
The entire information about the harmonic generation process is encoded in the
time-dependent dipole acceleration~\cite{BohanPRL98}. It is given by~\cite{NiikuraPRL05}
\begin{eqnarray}
  \label{eq:dipole_acceleration}
  \ddot{d}(t) = \langle\psi(t)|\nabla_x V(x)|\psi(t)\rangle\,, 
\end{eqnarray}
where the dependence on the external field is omitted since it does not
contain higher harmonics. The harmonic spectrum $S(\omega)$ is
obtained as~\cite{ConradJPhysB11} 
\begin{eqnarray}
  \label{eq:harmonic_spectrum}
  S(\omega) = |\ddot{d}(\omega)|^2 / \omega^2\,\,,
 \end{eqnarray}
where $\ddot{d}(\omega)$ is the Fourier transform of the dipole
acceleration~\eqref{eq:dipole_acceleration}. 

The electric field parameters given above lead to a ponderomotive
energy~\cite{BandraukPRA09} of $U_p = 0.16\,$Hartree such that 
the Keldysh adiabaticity
parameter~\cite{KeldyshJETP1965} becomes  $\gamma = 1.25$. 
Since with these parameters, $I_p>U_p>\omega_0$, where $I_p$ is the
ionization potential, the high harmonic generation process that we
consider procedes in the regime of above threshold
ionization (ATI)~\cite{Lewenstein2009}. Within the  quasi-classical
three-step model, the harmonic \textit{cutoff} position is given by~\cite{CorkumPRL93,LewensteinPRA94}
\begin{eqnarray}
  \omega_c &= & (I_p + 3.17U_p)/\omega_0\,.
\end{eqnarray}
For an electron in the ground state, it becomes  
$\omega_c=10.072$. The characteristic overestimation of the
recollision probability of 1D models with respect to their counterpart
3D models is here minimized by the choice of a few-cycle pulse~\cite{NiikuraPRL05}.

\subsection{Numerical performance}
\label{subsec:numerical_performances}

First, we compare the numerical performance of the multi-domain
pseudospectral method to that obtained with the mapped Fourier
grid. In both cases, we utilize the Chebychev propagator,
Eq.~\eqref{eq50}. For the multi-domain pseudospectral approach the
Hamiltonian is applied via sparse matrix-vector multiplications,
whereas the mapped Fourier grid method uses fast Fourier transforms
together with vector-vector multiplications in real and momentum
space.

We assume that initially the electron is in the ground state,
$|\varphi_0\rangle$, of the field-free Hamiltonian. In particular,
choosing $a=2$ in Eq.~\eqref{eq:potential} ensures that the ground
state energy coincides with that of the true Coulomb potential, namely
$0.5\,$Hartree.  
For the propagation based on the mapped Fourier grid, we use
$R_{max}=8000\,$ Bohr, which ensures that there are no spurious
reflections at the edges of the grid during propagation. The remaining parameters are chosen to yield fully converged results. 
Specifically, we find the grid to be converged  when using $2047$
coordinate points, which leads to a correct representation of
continuum states with  energies well above $E_{max} = 0.25\,$Hartree,
the highest continuum state that gets populated during the dynamics.

The eigenvalues and eigenfunctions are obtained by diagonalization of the
field-free Hamiltonian in the mapped Fourier grid representation. The set of
eigenvalues from $E_0$ to $E_{max}$ is used as a reference to define the
accuracy of the mapped pseudospectral method, when testing several
combinations of the number of domains, $M$, and collocation order $N$.
We find that for a low collocation order, which minimizes the 
spectral radius, a larger number of total points is needed than with 
the mapped Fourier grid. For example, $N_{pts} =2701$ for 
$M = 900$ and  $N=3$.

\begin{figure}[tb]
\begin{minipage}[tb]{\linewidth}
\centering 
\includegraphics[width=1.00\linewidth]{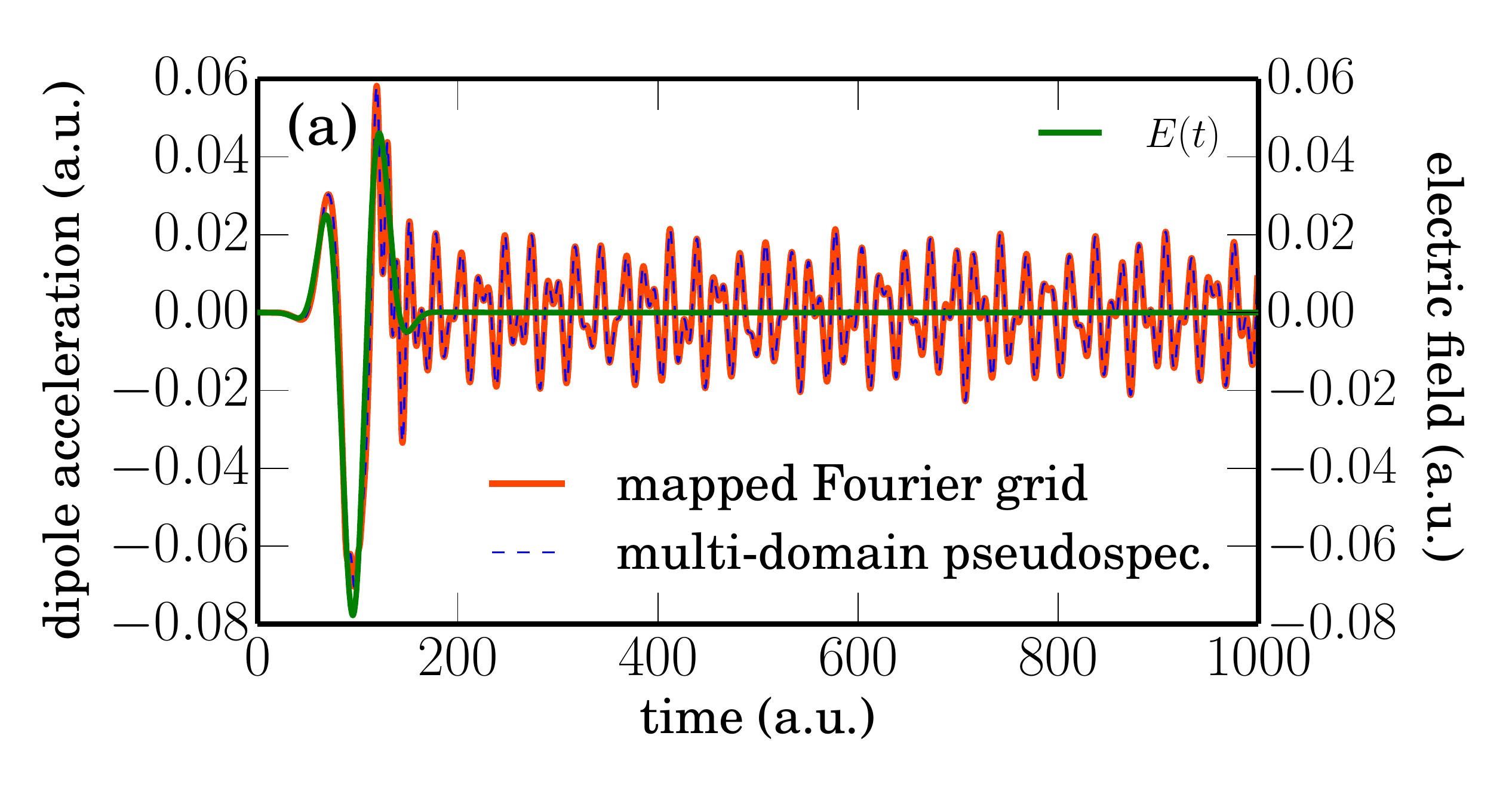}
\end{minipage} 
\begin{minipage}[tb]{\linewidth}
\centering 
\vspace{-0.4cm}
\includegraphics[width=1.00\linewidth]{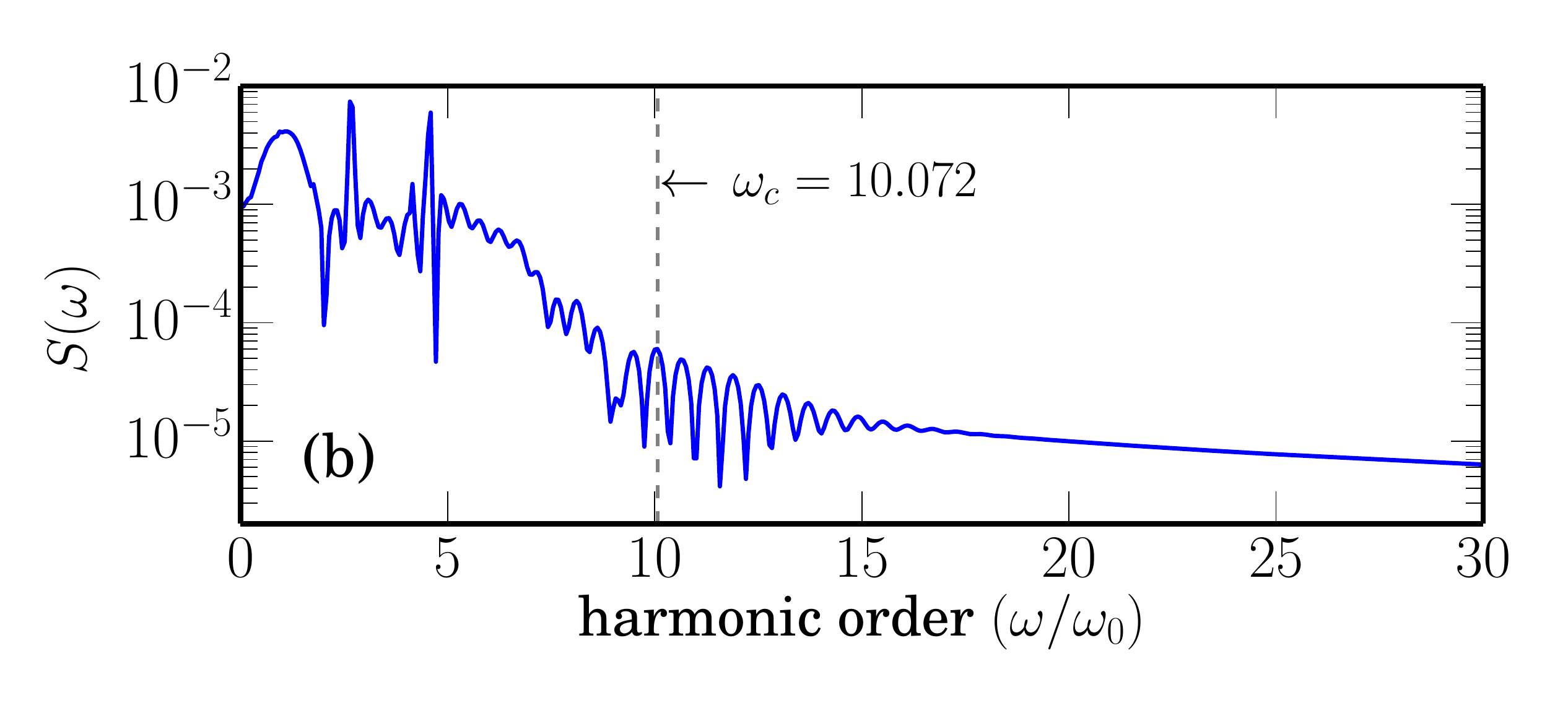}
\end{minipage}
\caption{(a) Time-dependent dipole acceleration $\ddot{d}(t)$
obtained with the
mapped Fourier grid method and the multi-domain pseudospectral
approach ($N=3$, $M=900$). For comparison, the electric field of
the driving pulse is also shown. 
(b)  Harmonic spectrum $S(\omega)$. 
} 
\label{fig:dipole_field}  
\end{figure}
\begin{table}[tb]
  \begin{tabular}{|c| c| c| c|}\hline
    $\quad N\quad$&$\quad M\quad$&spectral radius$^\dagger$  &CPU time$^\dagger$  \\\hline
    10    & 270& 973\% & 152\%    \\  
    6     & 450& 411\%  & 66\%    \\  
    5     & 540& 310\%  & 52\%    \\  
    4     & 675& 227\%  & 40\%    \\  
    3     & 900& 165\%  & 31\%    \\     \hline
    \multicolumn{4}{l}{$\dagger$\, relative to mapped Fourier grid method}\\
  \end{tabular}
  
  \caption{\label{tab:gll_efficiency} Numerical effort for wavepacket
    propagation with the adaptive-size multi-domain pseudospectral
    approach where $N$ denotes collocation order and $M$ the number of
    domains. The total number of collocation points is $N_{pts}=N\times M+1 =
    2701$. The reference calculation, using the mapped Fourier grid method and fast Fourier transforms,   with 2047 grid points and a spectral radius of
    1277.8$\,$Hartree., took 959$\,$s of CPU time.}
\end{table}
The dipole acceleration $\ddot{d}(t)$ and harmonic spectrum
$S(\omega)$ obtained with both propagation approaches are depicted in
Figs.~\ref{fig:dipole_field}(a) and (b), respectively. The few-cycle
laser pulse indeed induces a fast dynamics of the electron, and the
corresponding harmonic spectrum shows the characteristic cutoff. Clearly both
methods yield the same dynamics, as expected. The numerical
performance is, however, quite different. It is analyzed in
Table~\ref{tab:gll_efficiency}. Although the sparse structure of the
Hamiltonian matrix in the multi-domain pseudospectral approach leads
to a larger spectral radius, the CPU time required for propagation may
be  significantly smaller, depending on the collocation order $N$. Thus the multi-domain pseudospectral approach based on (sparse) matrix-vector multiplications is numerically more efficient than transforming the propagated wavepacked from coordinate to momentum representation by fast Fourier transforms, provided the parameters $N$ and $M$ are judiciously chosen. 

The role of the spectral radius becomes particularly apparent for the choice   $N=10$ and $M=270$ which leads to a propagation time 50 
per cent longer than that needed with mapped Fourier grid approach,
cf. Table~\ref{tab:gll_efficiency}. In this case, the spectral radius
is almost ten times larger than the one obtained with the mapped
Fourier grid. Correspondingly, the number of terms in the Chebychev
propagator, i.e., of applying the Hamiltonian, is ten-fold
increased. However, choosing $N=6$ and $M=450$ reduces the spectral
radius considerably, such that the CPU time for propagation is now
only two thirds of that using the mapped Fourier grid method. Already
for this choice of parameters, the adaptive-size multi-domain
pseudospectral approach starts to be more efficient. The efficiency
may be further improved by reducing $N$ and increasing $M$, up to a
third of the CPU time required with the mapped Fourier grid for $N=3$
and $M=900$. 

Note that the accuracy in all cases is roughly the same,
since the overall number of collocation points is sufficiently large.
A low collocation order $N$  minimizes the spectral radius, and thus the number of terms in the Chebychev propagator. Larger $N$ does not only lead to a larger spectral radius but also to a less sparse structure of the Hamiltonian, cf.~Fig.~\ref{fig:figure3.5}, i.e., it results in a two-fold increase in the numerical cost. Since small $N$ allows for highly accurate results, it is the preferrable choice. 
In summary, the best performance of the multi-domain pseudospectral approach is achieved by choosing a relatively large total number of points, with small $N$ and large $M$, such that the desired accuracy is obtained while minimizing the CPU time.

\subsection{Enhancement of the high harmonic yield}
\label{subsec:enhancement}

We now employ the time-dependent multi-domain pseudospectral approach
to analyzing the role of the initial state for the generation of the
harmonic spectrum, while keeping the driving pulse fixed (using the
same parameters as in Sec.~\ref{subsec:numerical_performances}). 
This perspective is different from earlier
studies~\cite{WerschnikJPB07,SchaeferPRA12,SolanpaaPRA14,ChouPRA15,CastroEuPhysJB15}
based on  optimal control theory which modified the
driving electric field to extend the harmonic cutoff and enhance the
harmonic yield. Specifically, we seek to answer the question whether
it is possible to enhance the yield of the harmonic spectrum at the
harmonic cutoff by a suitable preparation of the initial state. We
compare low-lying eigenstates of the field-free Hamiltonian as initial
state and superpositions thereof. These different initial states could
be prepared by a 'pre-pulse', preceding the pulse that drives the
harmonic generation. In contrast,
Refs.~\cite{SchaeferPRA12,SolanpaaPRA14,ChouPRA15,CastroEuPhysJB15}
only considered the ground state as initial state.

\begin{figure}[tb]
  \centering
  \includegraphics[width=1.00\linewidth]{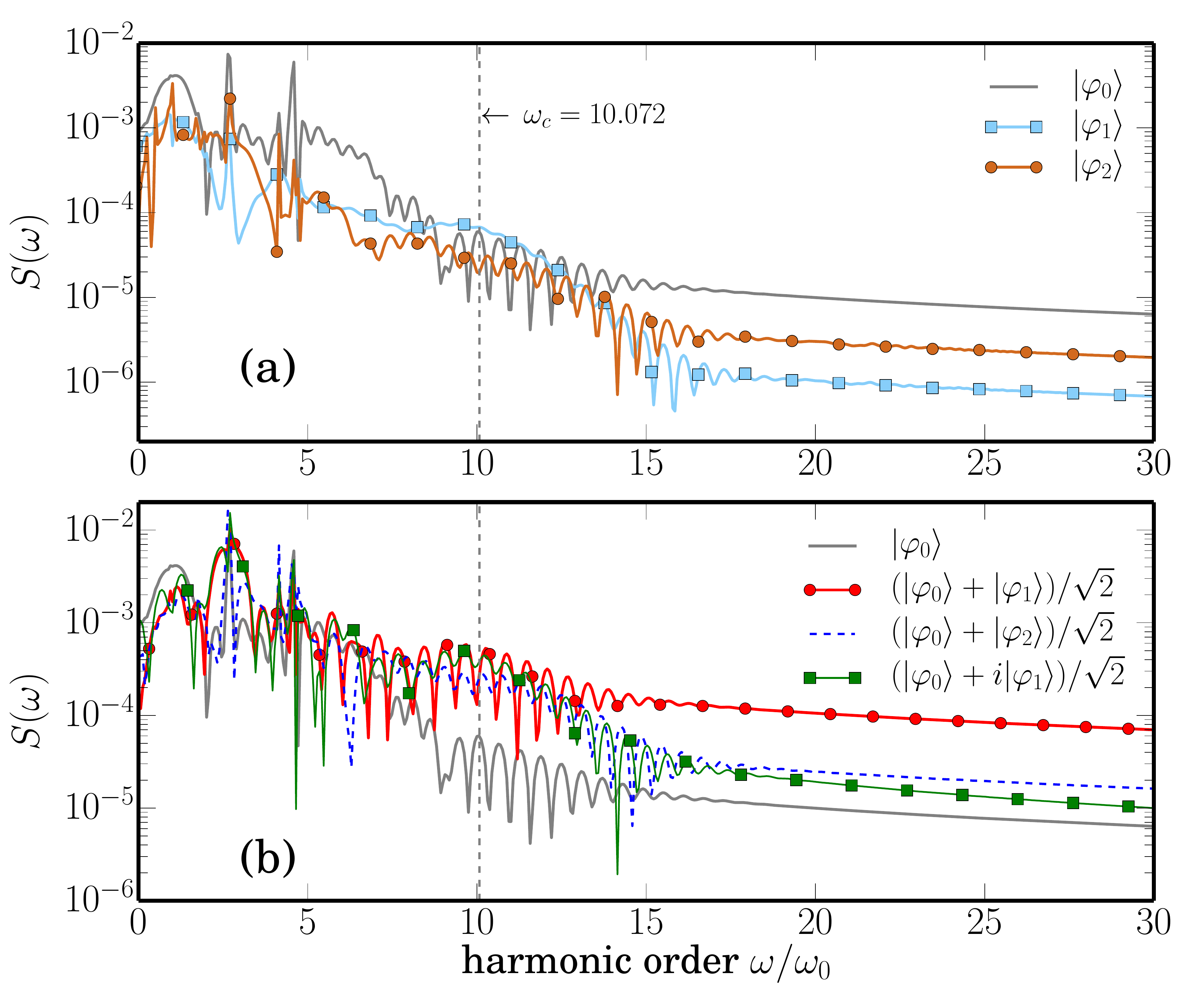}\vspace{-0.0cm}
  \caption{Harmonic spectrum for different initial states: (a)
    Eigenstates of the field-free Hamiltonian and (b) superpositions
    of two field-free eigenstates.}
  \label{fig:superposition}  
\end{figure}
Figure~\ref{fig:superposition}(a) shows the harmonic spectra obtained
for the first three eigenstates $|\varphi_0\rangle$, 
$|\varphi_1\rangle$ and $|\varphi_2\rangle$ of the field-free Hamiltonian, 
with eigenenergies $-0.500\,$Hartree, $-0.233\,$Hartree and
$-0.134\,$Hartree, as initial state. The exponential decay instead of
a plateau in Fig.~\ref{fig:superposition}(a) is characteristic of soft
core potentials, the plateau being attributed to the singularity
present in the Coulomb potential~\cite{GordonPRA05}. Since the
ionization potential is the largest for the ground state,
$|\varphi_0\rangle$ results in the largest harmonic cutoff, 
$\omega_c/\omega_0=10.1$ compared to $7.4$ and $6.4$ for the first
first and second excited state,
respectively. Figure~\ref{fig:superposition}(a) also shows that the
spectral yield of the high orders is largest for the ground state, 
particularly for higher photon energies.

Next, we consider, in Fig.~\ref{fig:superposition}(b), superpositions
of  field-free eigenstates as initial state and compare them to the
best single eigenstate, $|\varphi_0\rangle$. It is worth mentioning
that some precaution is necessary in the evaluation of the expectation
value~\eqref{eq:dipole_acceleration} since a superposition of
eigenstates leads to a dipole acceleration even without any driving
pulse. For instance, for a superposition of two
states, this ``field-free'' dipole acceleration is given by 
\begin{eqnarray}
  \label{eq:zero_field_acc}
  \ddot{d}_{ff}(t) &=&  2|a_i|\,|a_j| \cos(\omega_{i,j}\,t - \vartheta)\,
   \langle\varphi_i|\nabla_{x}V(x)|\varphi_j\rangle\,\,\quad\quad
\end{eqnarray}
where $\hbar\omega_{i,j}$ is the energy difference between the superimposed 
states, $\vartheta$ their relative phase and $|a_k|$ the norm of the expansion coefficients. 
In order to analyze true high harmonics, we focus on the spectral yield for frequencies well
above $\omega_{i,j}$, for example the yield close to the cutoff frequency. 

We consider an equal superposition of two field-free
eigenstates, allowing also for a complex
phase. Figure~\ref{fig:superposition}(b) reveals, that depending on 
the expansion coefficients in the initial state, the harmonic yields is
considerably enhanced, compared to the best single eigenstate,
$|\varphi_0\rangle$. The superposition $(|\varphi_0\rangle +
|\varphi_1\rangle)/\sqrt{2}$ does not only result in a higher harmonic
yield at the cutoff, but also in a larger integrated spectrum, i.e., a
larger integrated power density, for frequencies higher than
$\omega_c$. This is true not only for the comparison with the initial
states shown in Fig.~\ref{fig:superposition}(b), but also for other
superpositions. 

The finding of Fig.~\ref{fig:superposition}(b) motivates a more
thorough control study which is easily possible, given the numerical efficiency of the multi-domain pseudospectral approach.
Specifically, we use optimization to determine the best combination of
eigenstates, such that the power density of the harmonic yield
starting from the cutoff $\omega_c$ is maximized. This choice ensures
maximization of the total integrated spectrum for high harmonic orders
beyond the cutoff. In detail, we employ the Sequential PArametrization
(SPA) technique~\cite{GoetzPRASPA} to determine the expansion
coefficients in the initial state, $c_j\in\mathbb{C}$, such that
propagation of this state maximizes the integrated spectrum~\cite{SchaeferPRA12,SolanpaaPRA14}, 
\begin{eqnarray}
 \label{eq:Jw_functional}
    J[c_j] = \int_{\omega_c}^{\omega_f}|\ddot{d}(\omega)|^2 d\omega\,.
\end{eqnarray}
The harmonic \textit{cutoff} position $\omega_c$ is taken to be the
one obtained for the ground state as initial state. The upper limit is
defined to be $\omega_f = 3\omega_c$. Note that the functional as
defined in Eq.~\eqref{eq:Jw_functional} does not only
enhance the spectral yield in $[\omega_c, \omega_f]$, but it can also
extend the harmonic cutoff as a function of $\omega_f$.

We use $(|\varphi_0\rangle + |\varphi_1\rangle)/\sqrt{2}$ to start the
optimization, since this superposition  was found to considerably
enhance the power spectrum. The SPA technique updates the expansion
coefficients, which can take complex values, sequentially: Starting
with two guess coefficients, $c_0=c_1=1/\sqrt{2}$, additional
coefficients are sequentially added, once a plateau is encountered in the optimization~\cite{GoetzPRASPA}.

Upon optimization with only two states, we find the optimal initial
superposition to be composed of $|\varphi_0\rangle$ and $|\varphi_1\rangle$ with
coefficients $c_0 = 0.7215$ and $c_1=0.6924$. The resulting harmonic
yield is very slightly better, by less than 1 per cent, than that
obtained with equal weights, $c_0 =c_1\approx 0.7071$, in the initial superposition. A similarly small improvement is
obtained for a superposition involving $|\varphi_0\rangle$ and
$|\varphi_2\rangle$. Including up to eight eigenstates in the initial
superposition improves the value of the optimization
functional~\eqref{eq:Jw_functional} by $19$ per cent compared to
the superposition of $|\varphi_0\rangle$ and $|\varphi_1\rangle$ with equal weights. This improvement is, however, solely due to the smooth, exponentially
decaying region (data not shown), where the harmonic yield is already
small. 

\begin{figure}[tb]
\begin{minipage}[tb]{\linewidth}
\centering
\hspace{0.1cm}\includegraphics[width=0.91\linewidth]{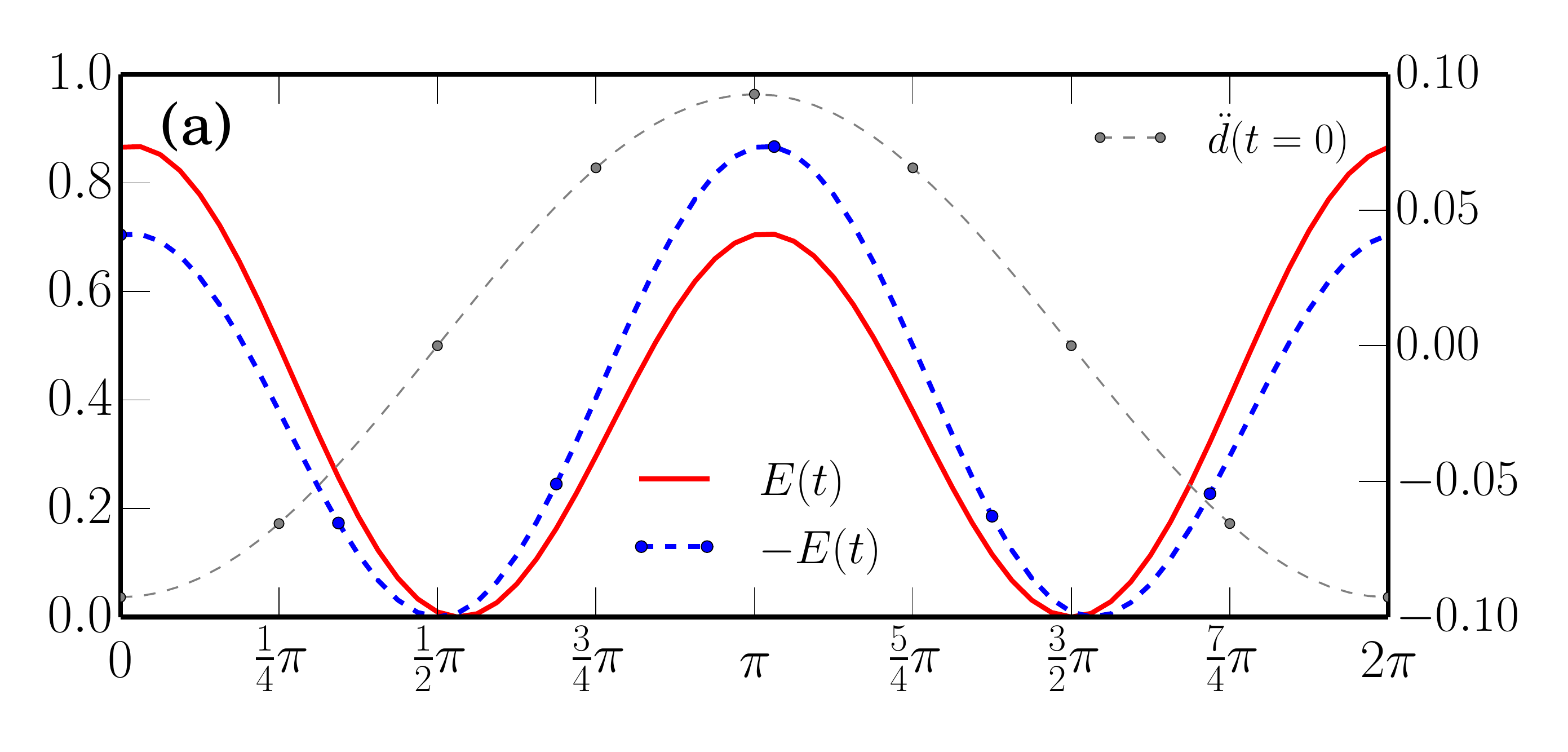}
\end{minipage} 
\begin{minipage}[tb]{\linewidth}
\centering
\vspace{-0.4cm}
\hspace{-0.1cm}\includegraphics[width=1.00\linewidth]{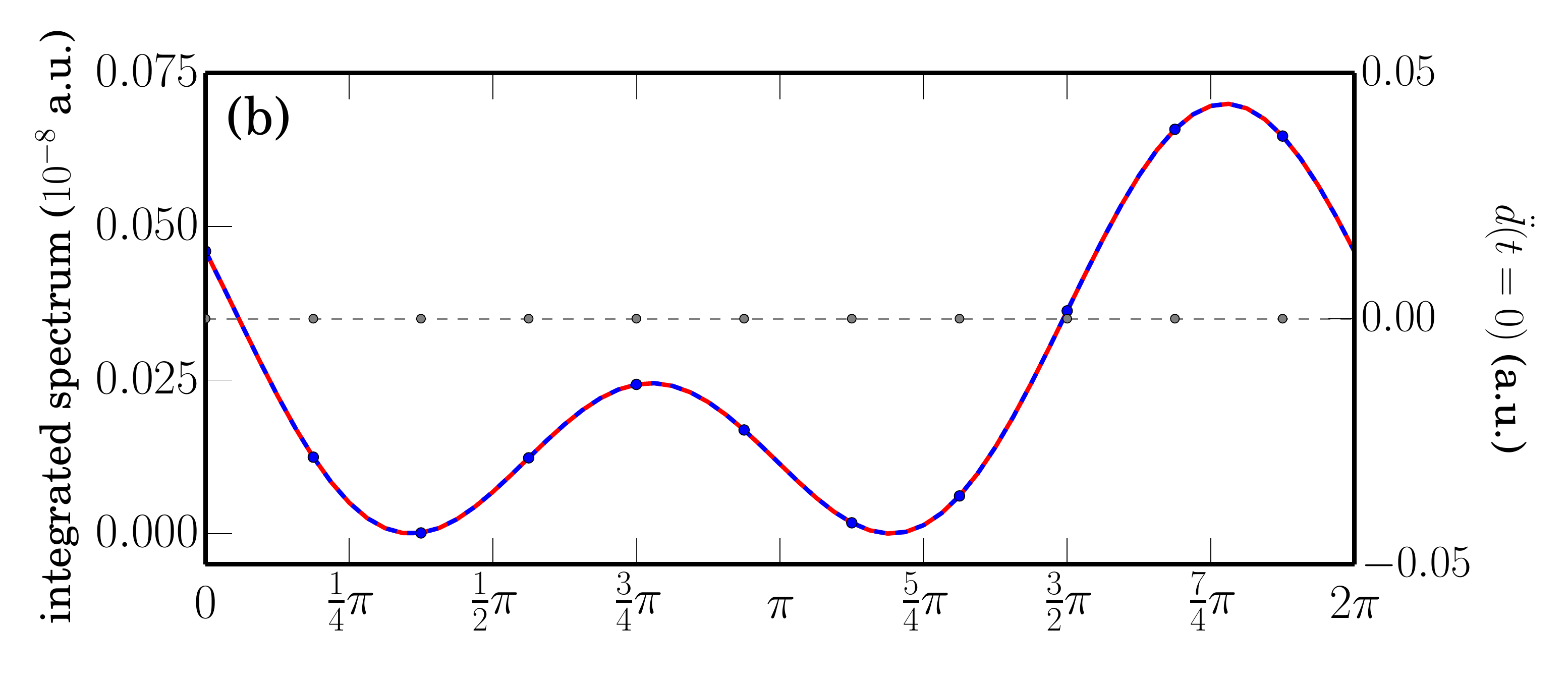} 
\end{minipage} 
\begin{minipage}[tb]{\linewidth}
\centering
\vspace{-0.4cm}
\includegraphics[width=0.93\linewidth]{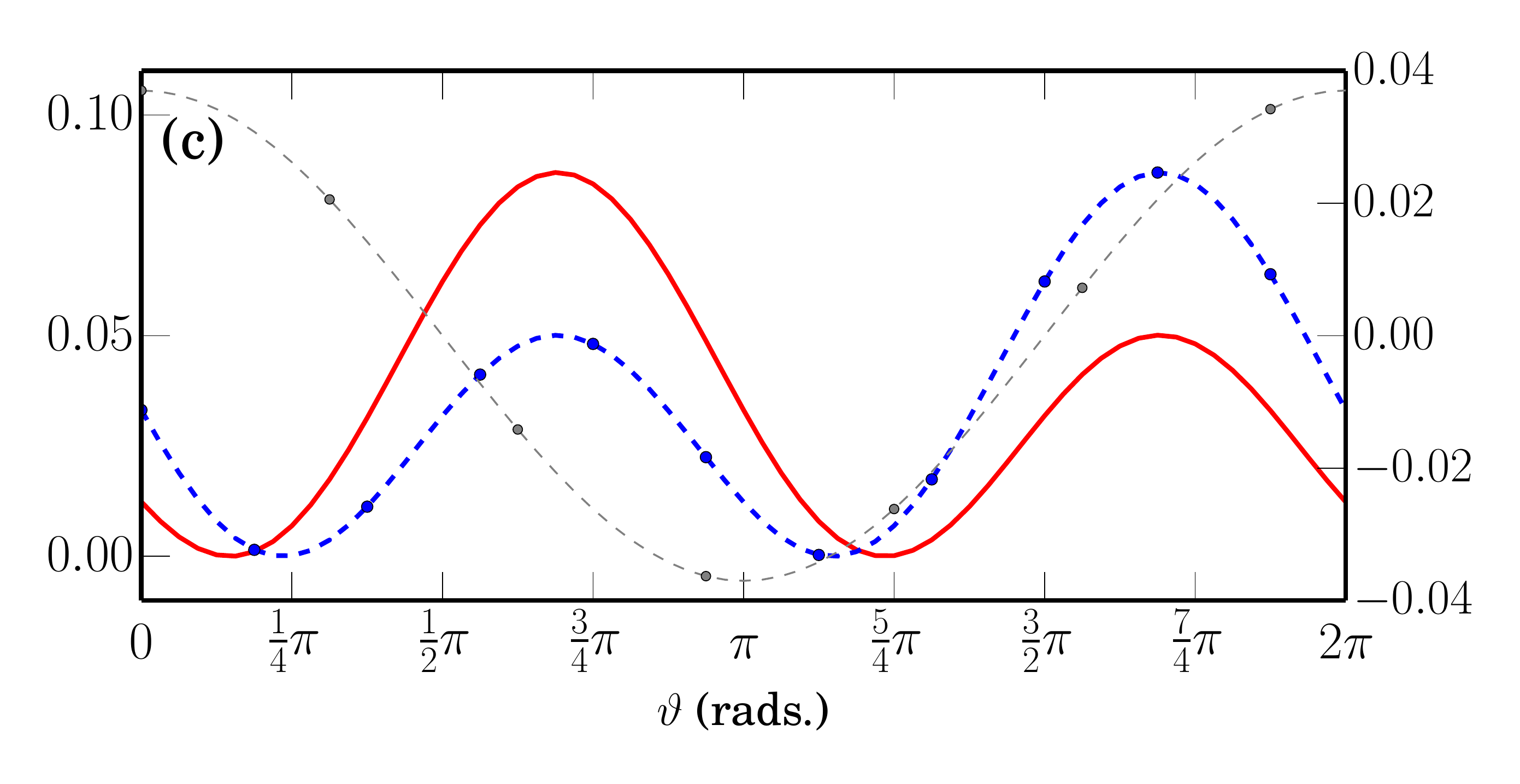}
 \caption{Integrated spectrum (red solid and blue dashed curve, left-hand side label), cf. Eq.~\eqref{eq:Jw_functional}, 
 and initial dipole acceleration (grey dotted curve, right-hand side label) as a function of the relative phase $\theta$  in Eq.~\eqref{eq:superposition_phase} for
 initial superpositions of $|\varphi_0\rangle$ and 
 $|\varphi_1\rangle$ (a), $|\varphi_0\rangle$ and $|\varphi_2\rangle$ (b) 
 and $|\varphi_0\rangle$ and $|\varphi_3\rangle$ (c) with electric
 fields $E(t)$ (red solid curve) and $-E(t)$ (blue dashed curve). 
 }
\label{fig:role_phase}  
\end{minipage}
\end{figure}
The role of the phase in the initial superposition is analyzed in Fig.~\ref{fig:role_phase}. It displays the integrated spectrum within 
the interval $[\omega_c,3\omega_c]$ as a function
of the relative phase $\vartheta$ in the superposition,
\begin{eqnarray}
  \label{eq:superposition_phase}
  |\varphi\rangle = \dfrac{1}{\sqrt{2}}(|\varphi_0\rangle
  + e^{i\vartheta}|\varphi_j\rangle)
\end{eqnarray}
for $j=1,2,3$. For $j=1$, maxima are found in Fig.~\ref{fig:role_phase}(a)
for $\vartheta\approx\pi/32$ and $\vartheta\approx
65\pi/64$ which result in the same maximal yield, differing from the yield for $\theta=0$, $2\pi$ by only $\approx 0.15$ per
cent. In contrast, the minimal yield observed in
Fig.~\ref{fig:role_phase}(a) differs by four orders of magnitude. 

In order to elucidate the physical origin of the oscillations of the harmonic yield as a function of the relative phase in the initial superposition state, we compare the integrated spectrum (solid red lines in Fig.~\ref{fig:role_phase}(a)) to the initial dipole acceleration
(dashed grey line) which is a direct result of the superposition,
cf. Eq.~\eqref{eq:zero_field_acc}. 
Indeed, the oscillations of the
spectral yield as a function of the superposition phase are strongly
correlated to the absolute value of the initial dipole acceleration
(grey dotted curve). Consider in particular the two initial states
$|\varphi\rangle = (|\varphi_0\rangle \pm
|\varphi_1\rangle)/\sqrt{2}$, i.e., $\vartheta = 0$ and $\vartheta =
\pi$. These states are orthogonal and lead to equal initial dipole
accelerations  with opposite sign but slightly different spectral
yields. 
This raises the question whether the sign of $\ddot{d}(t=0)$ determines the maximal value of the harmonic yield. In order to answer this question,  we compare 
the integrated spectrum obtained with $-E(t)$ to that for $E(t)$ (blue dashed and red solid lines in Fig.~\ref{fig:role_phase}(a)). The idea is that there is an effective ``initial'' time when the driving field starts to become non-zero. The superposition at $t=0$ prepares an ``initial'' wave packet at that time or, classically spoken, the dipole acceleration at $t=0$ determines the effective ``initial'' dipole acceleration at $t=t_p$. If the harmonic generation depends on both norm and sign of the dipole acceleration when the field starts to become non-vanishing, that is at $t=t_p$, a symmetric relationship should be found when changing the sign of  $E(t)$ at $t = t_p$. 
This symmetry is indeed observed in Fig.~\ref{fig:role_phase}(a), cf. the harmonic yield obtained with  $\vartheta = \pi$ (giving a positive 
$\ddot{d}(0)$) and $-E(t)$, which matches exactly 
the yield for  $\vartheta=0$ (giving a negative $\ddot{d}(0)$) and $+E(t)$. 
Shifting the electric field according to $E(t - T_s)$ with $T_s = 2\pi
/ \omega_{0,1}$ so that $\ddot{d}_{ff}(t) =\ddot{d}_{ff}(t-T_s)$ does not change 
the spectral yield (data not shown). This is of course expected for an initial condition at $t_p-T_s$ that is identical to that at $t_p$. 

To further investigate the dependence on the initial state, we
consider a superposition of eigenstates of the same parity, i.e., 
$|\varphi_0\rangle$ and $|\varphi_2\rangle$, cf. Fig.~\ref{fig:role_phase}(b). In fact, because $dV(x)/dx$ has odd parity, 
this superposition should lead to a vanishing initial dipole acceleration, cf. Eq.~\eqref{eq:zero_field_acc}. Therefore, the harmonic yield obtained with such an initial superposition should 
be not sensitive to a change of $E(t)$ to $-E(t)$, if the classical picture is still valid. 
This is indeed observed in Fig.~\ref{fig:role_phase}(b). Similarly, the superpositions with vanishing
initial dipole acceleration in Fig.~\ref{fig:role_phase}(a) are also
not sensitive to a change of $E(t)$ to $-E(t)$.  While for a superposition of $|\varphi_0\rangle$
and $|\varphi_1\rangle$, peaks in the high harmonic yield are found for
$\vartheta=0$ and $\vartheta = \pi$, i.e., for a  maximal initial dipole
acceleration (in absolute value), 
such a correlation is not observed for the superposition of $|\varphi_0\rangle$
and $|\varphi_2\rangle$. In this case, the dependence of the high
harmonic yield, for example the peak at $\vartheta=7\pi/4$, cannot be
explained based on a simple classical argument.

Figure~\ref{fig:role_phase}(c) displays another example of an initial superposition of even and odd parity states ($|\varphi_0\rangle$
and $|\varphi_3\rangle$). While a similar dependence on the sign of the
initial dipole acceleration is observed as in
Fig.~\ref{fig:role_phase}(a), in particular when changing the sign of 
the driving field, there is no one-to-one correlation between
the high harmonic yield and the initial dipole acceleration. This
shows that not only the initial dipole acceleration contributes to
an enhancement of the high harmonic   yield,  but it also depends on
the states involved in such a superposition.

\begin{figure}[tb]
 \includegraphics[width=1.00\linewidth]{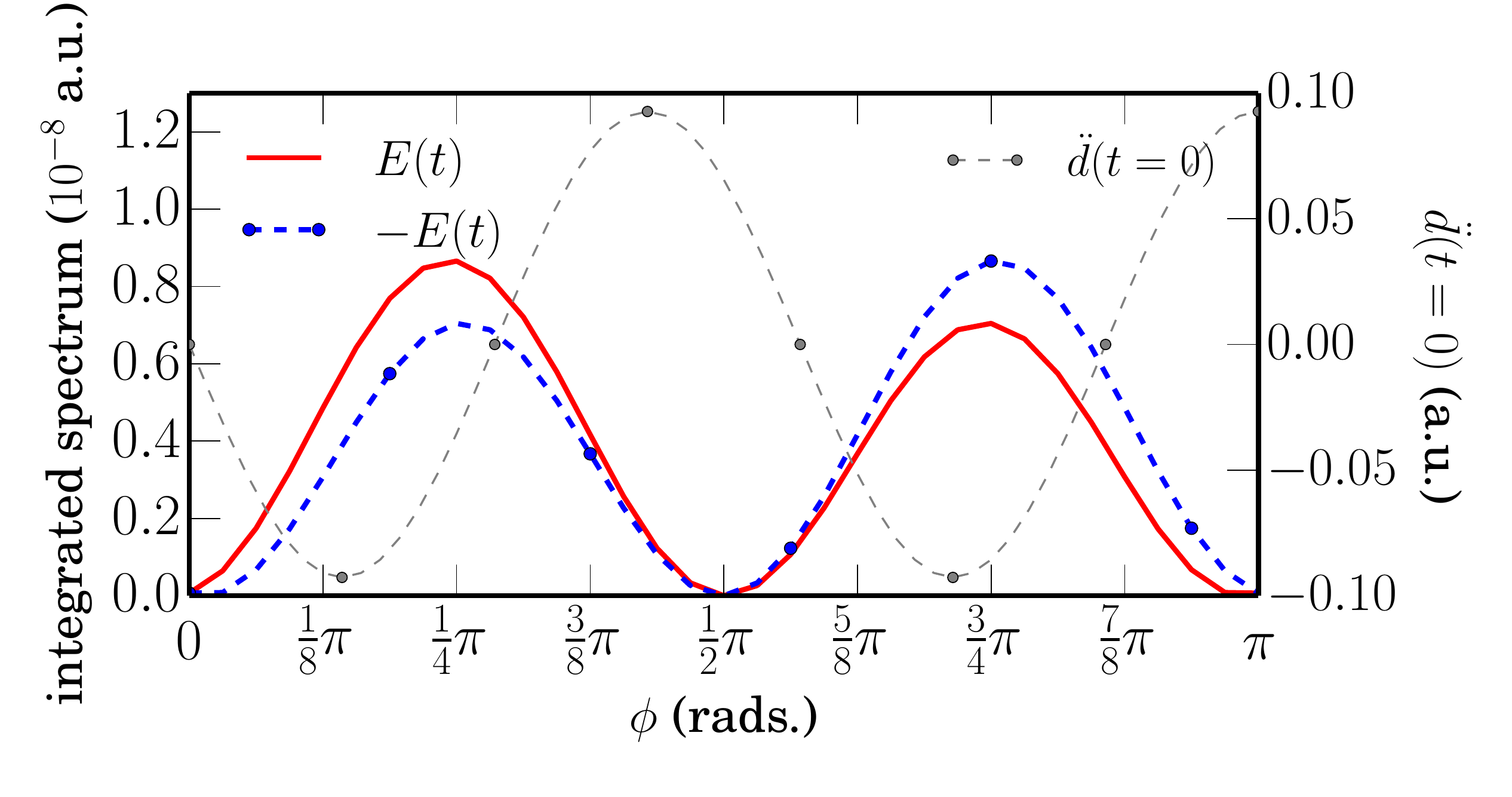}\vspace{-0.1cm}
\caption{Same as Fig.~\ref{fig:role_phase} but as a function of the
  relative amplitude of
  $|\varphi_0\rangle$ and $|\varphi_1\rangle$ in the initial
  superposition state, cf. Eq.~\eqref{eq:real_superposition_phase}. 
}
\label{fig:role_phase2}  
\end{figure}
Finally, we consider amplitude control of the initial superposition
state. This can be expressed as a function of a rotation angle $\phi$,
\begin{eqnarray}
\label{eq:real_superposition_phase}
|\varphi\rangle = \cos(\phi)|\varphi_0\rangle   
+ \sin(\phi)|\varphi_1\rangle\,\,.
\end{eqnarray}
The high harmonic yield as a function of  $\phi$, i.e., the relative
amplitude in a superposition of ground and first excited state is
shown in  Fig.~\ref{fig:role_phase2}. 
A correlation between the oscillations of the high harmonic yield and
the initial dipole acceleration is observed, similar to that found in
dependence on the relative phase. Also, an analogous symmetry when 
changing the sign of $E(t)$ is obtained. This shows that the control
over the high harmonic yield can equally be achieved by controlling
the relative phase or the relative amplitudes in the initial
superposition state.

\begin{figure}[tb]  
\centering
\vspace{-0.1cm}
\includegraphics[width=1.00\linewidth]{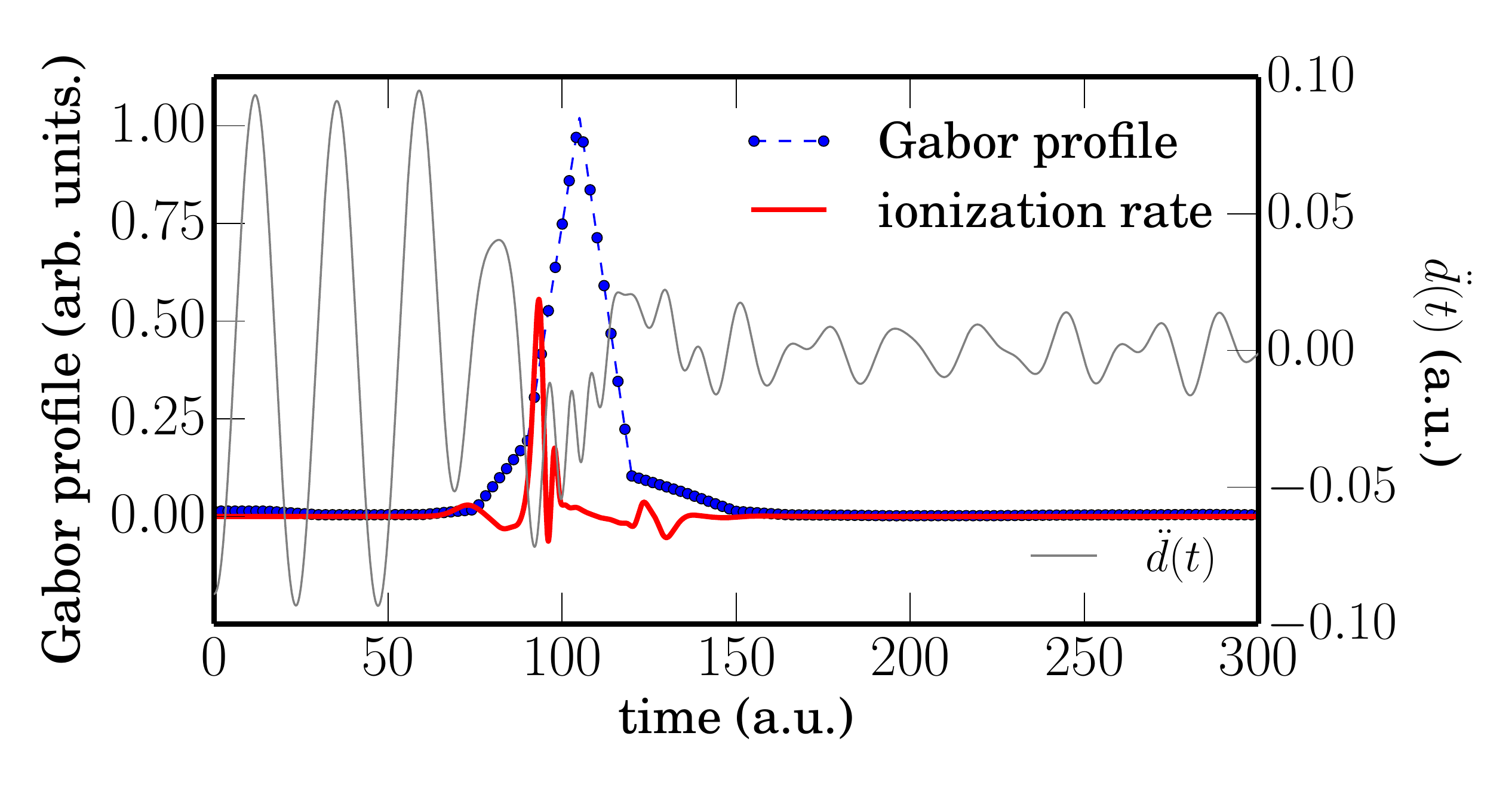}
\caption{Gabor profile of harmonics emitted with $\omega\in[\omega_c,3\omega]$ for the initial superposition state $|\varphi\rangle = (|\varphi_0\rangle
+ |\varphi_1\rangle)/\sqrt{2}$. The generation of the high harmonic coincides
with the temporal oscillations of the ionization rate, cf. full-red lines (scale not shown), in agreement with the three-step model. The grey line displays the dipole acceleration. 
}
\label{fig:Gabor_et_all}
\end{figure}
Enhancement of the high harmonic yield due to a purely quantum effect is
in contrast 
with the fact that high harmonic generation is usually explained with the 
three-step model~\cite{CorkumPRL93}, i.e., using a semi-classical picture. We 
therefore verify  whether the three-step model is still valid when starting from a 
superposition state. To this end, we plot in Fig.~\ref{fig:Gabor_et_all} the 
temporal Gabor profile of the harmonic yield corresponding to the frequency range 
above the cutoff and compare it to the ionization probability. The superposition 
of ground and first excited state, which results in the largest enhancement of 
the high harmonic yield, cf.~Fig.~\ref{fig:superposition}, is used as initial state. 
As can be seen from Fig.~\ref{fig:Gabor_et_all}, high harmonics are generated 
when the ionization probability is strongly time-dependent. The oscillations of 
the ionization probability (solid red line in Fig.~\ref{fig:Gabor_et_all}) indicate 
that the propagated wavefunction, or part of it, populates the continuum and 
then returns to the ionic core. This process of 
ionization and recombination is repeated several times. It is exactly in 
coincidence with the oscillations in the ionization probability that high 
harmonics are emitted as revealed by the peak in the Gabor profile 
(blue dashed line). We thus find the process of high harmonic generation still 
to be well described by the three-step model, in which semi-classical electron 
motion leads to the emission of high harmonics~\cite{CorkumPRL93}. This is in 
line with earlier findings that coherent control of high harmonic generation 
reduces to the problem of laser control over classical electron 
trajectories~\cite{LewensteinPRA94,SolanpaaPRA14}.

The dependence of the high harmonic yield on the relative phase in the
initial superposition points to constructive and destructive
interference in the maximization of the dipole acceleration, a
hallmark of coherent control. Controlling the harmonic yield by a
suitable preparation of the initial state could be realized in an
experiment with two pulses, a first pulse that prepares the desired
superposition state and a second pulse that drives the harmonic
generation. The time delay between the two pulses adjusts the relative
phase. To the best of our knowledge, such a strategy has not yet been
utilized for maximizing the yield at the cutoff in high harmonic
generation.

\section{Summary and Conclusions}
\label{sec:conclusion}

We have constructed a multi-domain pseudospectral representation of
the Hamiltonian and have employed it to solve the time-dependent
Schr\"odinger equation for the process of high order harmonic
generation. The advantage of our approach is that it allows for large
grids by adapting the size of each domain to the local kinetic
energy. Continuity between domains is ensured by employing Gauss-Lobatto
collocation and a weak formulation of the Schr\"odinger equation. The
resulting Hamiltonian matrix is sparse, yet the representation is
accurate due to accuracy of Gauss interpolation. When combined with
the Chebychev propagator for time evolution, it is important to keep
the spectral radius as small as possible. For a given desired
accuracy, this can be achieved by a judicious choice of the number of
domains and collocation order. For the example of high harmonic
generation for an electron subject to a soft Coulomb potential, we
have found our approach to be faster than the mapped Fourier grid
method by a factor of about three. 

The advantage of our approach is its stability and accuracy, besides
efficiency. These features derive from the pseudospectral treatment of
both spatial degree of freedom and time
dependence~\cite{KosloffJPC88,RonnieReview94}. Our approach is thus
particularly suitable for problems where a large grid and long
propagation times are needed, for example to calculate spectra in
photoionization. It can also be employed in multi-dimensional problems
where the sparsity of the Hamiltonian representation will be even more important.

Efficient and accurate propagation methods are also a prerequisite in
optimal control studies~\cite{GlaserEPJD15} where iterative algorithms
require many
propagations to maximize the figure of merit. We have benefited from
the efficiency of the  multi-domain pseudospectral representation of
the Hamiltonian combined with Chebychev propagation to maximize the
yield of high order harmonics. In particular, we have found that an
initial superposition state may significantly enhance the integrated
high harmonic power density. This is complementary to recent demonstrations of coherent control of high harmonic generation that have exploited high lying electronically excited states~\cite{BeaulieuPRL16} and nuclear motion~\cite{LaraPRL16}. In our control scheme, 
superimposing the lowest two eigenstates
with equal weights improves the harmonic yield at the so-called cutoff
frequency by one order of magnitude. The relative phase in the initial
superposition is found to be important, pointing to a coherent control
mechanism for the harmonic yield. Such a control could be realized by
a pre-pulse to prepare the initial superposition state and proper
choice of the time delay of the pulse driving the harmonic
generation.

\begin{acknowledgments}
  Financial support by the State Hessen Initiative for the
  Development of Scientific and Economic Excellence (LOEWE) within the
  focus project Electron Dynamic of Chiral Systems (ELCH) is                
  gratefully acknowledged. This research was supported in part by the National Science Foundation under Grant No. NSF PHY11-25915.
  A.S. gladly  acknowledges support by the Agence nationale de la recherche (contract No. ANR-12-BS04-0020).
\end{acknowledgments}

\appendix
\section{Collocation with Legendre polynomials}
\label{sec:Legendre}

The Legendre polynomials\cite{milton,boyd}
are the solutions of the second order differential equation
\begin{eqnarray}
\label{eq0}
\left.\left.\dfrac{}{}\right( (1-\xi^2) L^\prime_n(\xi) \right)^\prime + n(n+1)L_n(\xi) = 0\,,
\end{eqnarray}
where $^\prime$ denotes the first derivative with respect to the argument of $L_n(\cdot)$.  In the interval $\Lambda = [-1,1]$, the Legendre polynomials are orthogonal with respect to the L$_2$ inner product 
and they obey the three-term recurrence relation
\begin{eqnarray}
\label{eq1}
(n+1)L_{n+1}(\xi) = (2n+1)\xi L_n(\xi) - nL_{n-1}(\xi),\ n\ge 1\,,\nonumber\\
\end{eqnarray}
with $L_0=1$,  $L_1=\xi$, where $\xi\in\Lambda$. Another useful recurrence relation reads~\cite{milton}
\begin{eqnarray}
\label{eq2}
(2n+1)L_{n}(\xi) = L^{\prime}_n(\xi) - L^{\prime}_{n-1}(\xi),\ n\ge 1\,.
\end{eqnarray}

In the interval $\Lambda = [-1,1]$,\ the set 
$\{\xi_j,\omega^{\Lambda}_j\}$ is defined as the set of Gauss-Lobatto-Legendre
nodes $\xi_j$ and Gaussian quadrature weights $w^{\Lambda}_j$. It is given by~\cite{boyd}
\begin{eqnarray}
\label{eq3}
\left\{
\begin{array}{lll}
\{\xi_j\}_{0 \le j \le N } & = & \text{zeros of}\hspace{0.3cm} \zeta(\xi)= (1-\xi^2) L^\prime_N(\xi) \\ \\ 
\omega^{\Lambda}_j& = &  \dfrac{2}{N\ (N-1) (L_N(\xi_j))^2}\, .  \\ 
\end{array}
\right.
\end{eqnarray}

For moderate order collocation, the $N-1$ interior points of the Gauss-Lobatto-Legendre grid in $\Lambda = [-1,1]$ can be generated
with the help of the Golub-Welsh algorithm~\cite{GolubBook,GolubMathComp69}. 
In detail, using Eqs.~\eqref{eq1}-\eqref{eq3}
it is straightforward to find the recursion relation for $L_n(\xi)$,
\begin{subequations}
  \begin{eqnarray}
    \beta_{n} L^\prime_{n+1}(\xi) + \alpha_{n} L^\prime_{n-1}(\xi) - \xi L^\prime_{n}(\xi) = 0 \,,
  \end{eqnarray}
where the recursion coefficients $\alpha_n$ and $\beta_n$ are given by 
\begin{eqnarray}
  \alpha_n = \dfrac{n+1}{2n+1}\hspace{1.0cm}\text{and}\hspace{1.0cm}\beta_n
  = \dfrac{n}{2n+1}
\end{eqnarray}
Taking into account  Eq.~\eqref{eq3}, i.e., $L^\prime_{N}(\xi_j)=0$ for all $j=1,\dots,N-1$,  the  tridiagonal Jacobian matrix reads
\begin{widetext}
\begin{eqnarray}
\begin{pmatrix}
  0   & \beta_1   &0   &0     & \dots & 0 \\
  \alpha_2 & 0     &\beta_2 &0     & \dots & 0  \\
 \vdots & \ddots   & 0   & \ddots     &       & 0\\
  0     & \dots   & \alpha_{n}  &0      &\beta_{n}   & 0 \\
 \vdots &  0  & \dots   &  \ddots    &   \ddots    & 0 \\ 
  0     & \dots   & \dots  & 0      &\alpha_{N-1}   & 0 \\
\end{pmatrix}
\begin{pmatrix}
   L^\prime_1(\xi_j)\\ 
   L^\prime_2(\xi_j)\\ 
   \vdots\\
   L^\prime_n(\xi_j)\\
   \vdots\\
   L^\prime_{N-1}(\xi_j) \end{pmatrix}
    =  \xi_j\begin{pmatrix}L^\prime_1(\xi_j)\\
      L^\prime_2(\xi_j)\\
   \vdots\\
   L^\prime_n(\xi_j)\\
   \vdots\\
   L^\prime_{N-1}(\xi_j)\end{pmatrix}\,,\nonumber\\
\end{eqnarray}
\end{widetext}
\end{subequations}
where the eigenvalues correspond to the $N-1$ roots of $\L_N^\prime(\xi_j)$ which
define, according to Eq.~\eqref{eq3}, the interior points of the Gauss-Lobatto grid. The extrema are
given by $\xi_0 = -1$ and $\xi_N=1$. Alternatively, in particular for a high-order quadrature, it is suitable to use a Newton-root-finding iterative method in order to avoid round-off errors that may occur during the diagonalization of the Jacobian matrix.

A first order Taylor expansion of $\zeta(\xi)$, defined in Eq.~\eqref{eq3}, around the $j$th Gauss-Lobatto-Legendre point, i.e., the $j$th root of $\zeta(\xi)$, gives
\begin{eqnarray}
\label{eq13}
\begin{array}{lll}
  \zeta(\xi) & \simeq & \zeta(\xi_j) + \zeta^\prime(\xi_j)(\xi-\xi_j) + \mathcal{O}(\vert \xi-\xi_j\vert)^2\vspace{0.30cm}\\
& = &\zeta^\prime(\xi_j)(\xi - \xi_j)\,,
\end{array} 
\end{eqnarray}
since, by definition, $\zeta(\xi)$ vanishes at the Gauss-Lobatto-Legendre points, $\zeta(\xi_j)=0$.
Equations~\eqref{eq3} and~\eqref{eq13} yield an explicit expression of the Legendre cardinal function $\delta^{\Lambda}(\xi-\xi_j)$,
\begin{eqnarray*}
\delta^{(\Lambda)}(\xi-\xi_j) &=& \frac{\zeta(\xi)}{\zeta^\prime(\xi_j)(\xi-\xi_j)}
= \frac{L^\prime_N(\xi)(1-\xi^2)}{(L^\prime_j(\xi)(1-\xi^2))^\prime}\dfrac{1}{(\xi-\xi_j)}\,,
\end{eqnarray*}
where $\zeta(\xi)$ is defined in Eq.~\eqref{eq3} and $L_j$ denotes the $j$th Legendre polynomial. Together with Eq.~\eqref{eq0}, this yields
\begin{eqnarray}
\label{eq16}
\delta^{(\Lambda)}(\xi-\xi_j)\equiv - \dfrac{ L^\prime_N(\xi)(1-\xi^2)}{N(N+1)L_N(\xi_j)}\dfrac{1}{\xi-\xi_j}
\end{eqnarray}
Moreover, we have $\delta^{(\Lambda)}(\xi_i-\xi_j) = \delta_{ij}$ at each 
$\xi_j$ by construction which results in the first order differentiation matrix for Legendre cardinal functions, cf. Eq.~\eqref{eq17}. 


%

\end{document}